\documentclass[useAMS,usenatbib,fleqn]{mnras}

\usepackage{natbib}
\usepackage{subfig}
\usepackage{graphicx}
\usepackage{etex}
\usepackage{amsmath}
\usepackage{amssymb}
\usepackage[titletoc,title]{appendix}
\usepackage{hyperref}
\hypersetup{colorlinks=true,linkcolor=blue,citecolor=blue,filecolor=blue,urlcolor=blue}
\newcommand{\angstrom}{\textup{\AA}}
\usepackage{float}
\usepackage{xfrac}
\usepackage[T1]{fontenc} 
\usepackage{aecompl}

\title[]{Stellar populations, stellar masses and the formation of galaxy bulges and discs at $z < 3$ in CANDELS}
\author[Margalef-Bentabol et al.]{Berta Margalef-Bentabol$^{1}$\thanks{Berta.Margalef@nottingham.ac.uk}, Christopher J. Conselice$^{1}$, Alice Mortlock$^{2}$, \newauthor Will Hartley$^{3}$,  Kenneth Duncan$^{4}$, Rebecca Kennedy$^{1}$, Dale D. Kocevski$^{5}$,\newauthor Guenther Hasinger$^{6}$\\
$^{1}$School of Physics and Astronomy, University of Nottingham, Nottingham, NG7 2RD, UK\\
$^{2}$SUPA Institute for Astronomy, University of Edinburgh, Royal Observatory, Edinburgh EH9 3HJ, UK\\
$^{3}$Department of Physics and Astronomy, University College London, London, WC1E 6BT, UK\\
$^{4}$Leiden Observatory, Leiden University, 2300 RA Leiden, The Netherlands\\
$^{5}$Department of Physics and Astronomy, Colby Colege, Waterville, ME 04901, USA\\
$^{6}$Institute for Astronomy, 2680 Woodlawn Drive, University of Hawaii, Honolulu, HI 96822, USA
}

\date{Accepted XXX. Received YYY; in original form ZZZ}

\pubyear{2017}

\begin{document}
\label{firstpage}
\pagerange{\pageref{firstpage}--\pageref{lastpage}}
\maketitle

\begin{abstract}

We present a multi-component structural analysis of the internal structure of $1074$ high redshift massive galaxies at $1<z<3$ from the CANDELS HST Survey. In particular we examine galaxies best-fit by two structural components, and thus likely forming discs and bulges. We examine the stellar mass, star formation rates, and colours of both the inner `bulge' and outer `disc' components for these systems using SED information from the resolved ACS+WFC3 HST imaging. We find that the majority of both inner and outer components lie in the star-forming region of UVJ space ($68$ and $90$ per cent respectively). However, the inner portions, or the likely forming bulges, are dominated by dusty star formation. Furthermore, we show that the outer components of these systems have a higher star formation rate than their inner regions, and the ratio of star formation rate between `disc' and `bulge' increases at lower redshifts. Despite the higher star formation rate of the outer component, the stellar mass ratio of inner to outer component remains constant through this epoch. This suggests that  there is mass transfer from the outer to inner components for typical two component forming systems, thus building bulges from disks. Finally, using Chandra data we find that the presence of an AGN is more common in both $1$-component spheroid-like galaxies and $2$-component systems ($13\pm3$ and $11\pm2$ per cent) than in $1$-component disc-like galaxies ($3\pm1$ per cent), demonstrating that the formation of a central inner-component likely triggers the formation of central massive black holes in these galaxies.

\end{abstract}


\section{Introduction}\label{sec.intro}

In the local Universe galaxies have a well defined structure and can be mostly classified  into disc-dominated, spheroidal systems, or mixed systems with a spheroidal component surrounded by a disc. On the other hand, at higher redshift ($z>2$), the majority of galaxies are  peculiar \citep{Conselice05,Mortlock13,Huertas16}. Thus, during the epoch of $1<z<3$ galaxies must have undergone a large amount of evolution  resulting in, or produced by, morphological transformations. By $z=1$ the dominant galaxy population are systems that have an inner and an outer component (or bulge and disc) \citep{Bruce12, Mortlock13, Margalef16}. However, it is still unclear how and when these different types of galaxies form, and how they are related to each other. For example, do bulges form from instabilities in discs, or do bulges form first and then later accrete a disc?

There are in fact many different theories for the formation of bulges and discs. For example some models predict that bulges form at high redshift due to disc instabilities and grow via minor mergers \citep{Hopkins10}. Disc formation may also occur due to gas accretion around a bulge which formed early in the history of the Universe, either through a rapid collapse or from multiple mergers of galaxies. It is  thus still unclear which component forms first, and how these inner and outer components evolve together or separately. To solve these problems it is vital to investigate the internal properties of galaxies which have multiple components at high redshift and to trace the evolution of their properties through cosmic time.

The bi-modality observed in morphology at low redshift in the galaxy population appears to translate to star formation activity, with discs being blue and star-forming, and bulges and spheroids red and passive \citep{Strateva01, Kauffmann03}. This bi-modality in morphology and star formation activity is also observed at higher redshift \citep{Brammer09, Whitaker11}, but the separation is less clear, and there is some evidence for a population of blue bulges and red discs \citep{Bamford09,Conselice11}. Therefore a key question is how morphology and star formation/quenching are related. Different studies have proposed several mechanisms to explain this, such as morphological quenching \citep[e.g.,][]{Martig09}, in which a galaxy shuts down its star formation activity due to a morphological transformation. Another possible internal cause of galaxy quenching is the presence of an active galactic nuclei (AGN), in which case AGN feedback heats the surrounding medium, and thus prevents the accretion of cold gas \citep{Croton06, Bluck11}. However, there are other properties that may play an important role in the star formation activity or quenching of a galaxy, such as environment and stellar mass.

Traditionally galaxies at high redshift are studied as single component systems \citep{Buitrago08, Trujillo07}, due to the difficulty of resolving inner and outer (or bulge and disc)  components at high redshift. Thanks to the high-resolution of WFC3 \textit{HST}-images it is just possible to separate the surface brightness profiles of high redshift galaxies into their bulge and disc components \citep{Bruce12,Lang14}. A major problem is however distinguishing galaxies which are intrinsically $1$-component systems from those which are in a bulge and a disc \citep{Margalef16}. 

In this paper we compare the formation histories of different types of galaxies, as measured through their surface brightness profile shapes, and for those systems which can be decomposed into two components we examine the evolution of  inner and outer components as a function of redshift. We also explore how the rest-frame $U-V$ colour change with redshift, and the role of stellar mass in producing evolution within separate components. Finally, we investigate how the star formation activity of both the bulge and the disc evolves with cosmic time, and how this affects the assembly of stellar mass in galaxies.

With the CANDELS data from WFC3 and ACS we are now able to probe the components of galaxies from the ultra-violet to the near-infrared.
Studying galaxies and their components at theses different wavelengths allows us to probe the properties of bulges and discs separately. We do this by modelling the SEDs of the bulge and the disc, uniquely and distinctly, after decomposing the galaxy at all the observed wavelengths. From this, we are able to trace the evolution of the stellar populations within these components, as well as measure the contribution to the total stellar mass and star formation rate (SFR) from each of the components \citep{Bruce14a,Bruce14b}.

In addition to evolving in morphology from peculiar to Hubble type systems we know that galaxies become more massive as time progresses. There are different ways in which a galaxy can add stellar mass, such as through star formation bursts, or mergers, but it is not clear how the inner and outer components of galaxies grown together. Two component systems seem to add mass in the outer parts of the galaxy while they grow in size \citep{Margalef16}. However, the detailed evolution of mass within these components is still uncertain, and this is another issue we examine in this paper.

The structure of this paper is as follow. Section \ref{sec.data} is devoted to describing the data we use. In Section \ref{sec.method} we describe how we calculate the structural parameters, stellar masses, rest-frame colours and SFRs. In Section \ref{sec.results} the main results of the paper are gathered, and in Section \ref{sec.discussion} we discuss and summarize the results. Finally, an Appendix is included where we discuss simulations to determine the reliability of our results. Throughout this paper we use $AB$ magnitude units and assume the following cosmology: $H_0=70\mathrm{\ Km s}^{-1}\mathrm{ Mpc}^{-1}$, $\Omega_{\lambda}=0.7$, and  $\Omega_m=0.3$.

\section{Data}\label{sec.data}

\subsection{Imaging}\label{subsec.imaging}

For this work a sample of $1074$ galaxies from the CANDELS UDS field is selected, at redshifts $1< z < 3$ and with stellar masses $M_{\ast}\geq10^{10}\textrm{M}_{\odot}$. CANDELS \citep{Grogin11,Koekemoer11} is a Multi Cycle Treasury Program which images the distant Universe with two cameras on the \textit{Hubble Space Telescope}, the Wide Field Camera $3$ (WFC3) and the Advanced Camera for Surveys (ACS), and it covers an area of $800\ \textrm{arcmin}^2$ in five different fields: GOODS-N, GOODS-S, EGS, UDS and COSMOS. For this work we use data from the CANDELS UDS field in the \textit{V}, \textit{i}, \textit{J} and \textit{H} bands, within the region where WFC3 and ACS overlap, which comprise an area of $187\ \textrm{arcmin}^2$. The $5\!\,\sigma$ point-source sensitivities in $AB$ magnitudes for this filters are $H=27.1$, $J=27.0$, $i=28.4$ and $V=28.4$.

The CANDELS UDS field is complemented by a large number of multiwavelengh observation of the larger UDS region, including: \textit{B}, \textit{V}, \textit{R}, \textit{i}, \textit{z}-band data from SXDS, \textit{U}-band CHFT data, \textit{J}, \textit{H} and \textit{K}-band data from UKIDSS,  $F606W$ and $F814W$-band \textit{HST} ACS data, $H160$ and  $J125$-band \textit{HST} WFC3 data and \textit{Y} and \textit{Ks} band data from HAWK-I UDS and GOODS-S survey (HUGS; VLT large program ID 186.A-0898, PI: Fontana; \citealt{Fontana14}). Additionally, for CANDELS UDS there is IRAC channel $1$ and $2$ ($3.6$ and $4.5$ $\mu \textrm{m}$) data from the \textit{Spitzer} Extended Deep Survey (SEDS; PI: Fazio; \citealt{Ashby13}). For further discussion of the CANDELS UDS region photometry see \cite{Galametz13}.

\subsection{Sample of Galaxies}

We use a sample of galaxies from \cite{Margalef16}. Redshifts and initial stellar masses are computed as described in \cite{Mortlock15} and \cite{Hartley13}. This sample is selected from the WFC3 region of the UDS, and comprises galaxies at redshifts $1<z<3$ and $M_{\ast}>10^{10}M_{\odot}$. For this work, in addition to near infrared images ($H$-band and $J$-band) from the WFC3 camera, we also use the $V$-band imaging and $i$-band from the ACS camera. The regions that these two cameras map do not completely overlap, and therefore the final sample for this work ($1074$ galaxies) only consists of galaxies with both near-infrared and visible-light images.
 
\begin{figure}
  \includegraphics[width=1\linewidth]{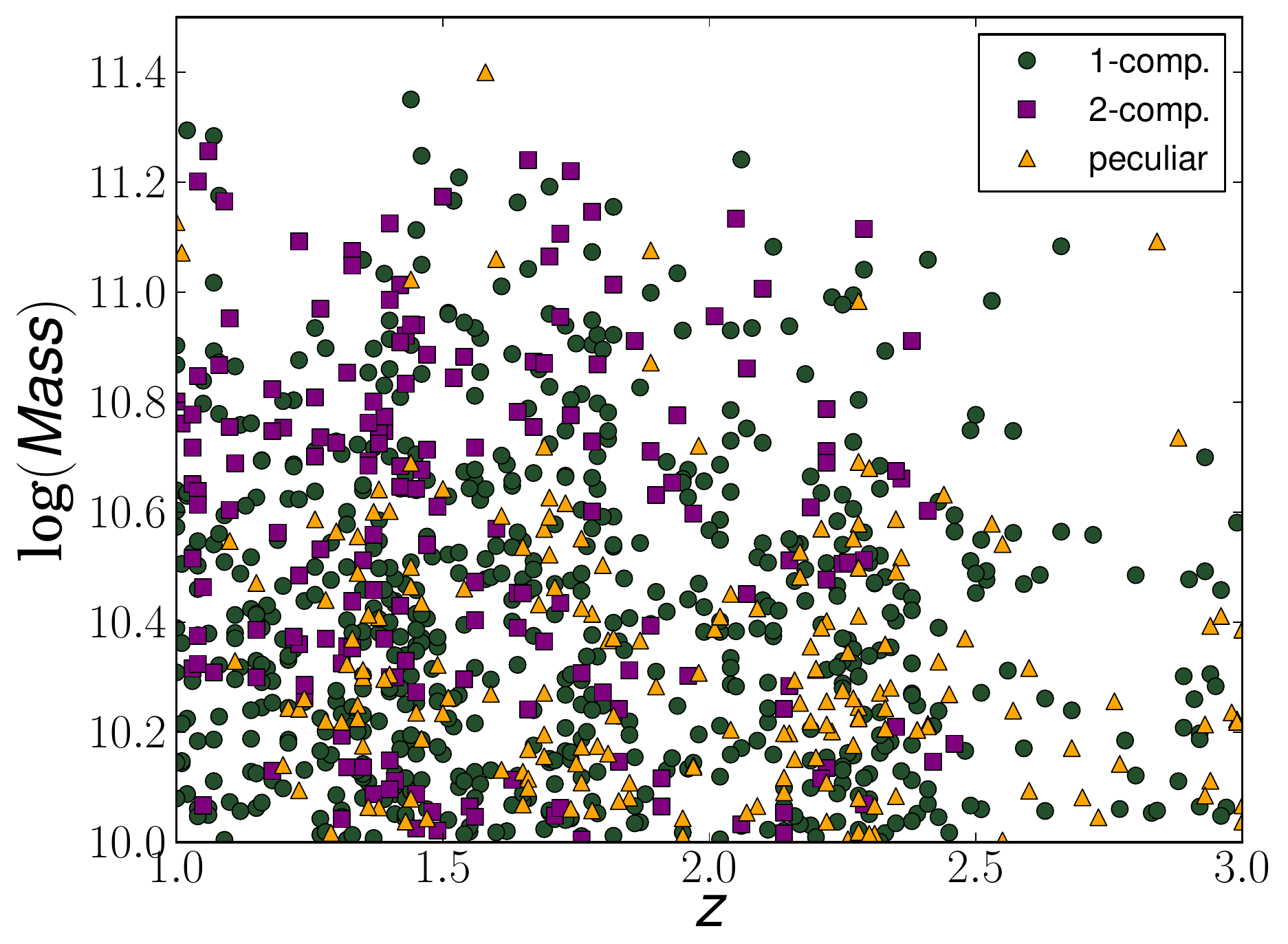}
  \caption{The stellar mass distribution with redshift for our initial stellar mass selected sample of galaxies (as calculated in \protect\citealt{Mortlock15}), for $1$-component galaxies (green circles), $2$-component galaxies (purple squares) and peculiars (yellow triangles).}\label{fig.sample}
\end{figure}

Galaxies in our sample are classified as either $1$-component, $2$-component or peculiar galaxies. The classification is done following the method in \cite{Margalef16}, in which different methods (Visual classification, $F$-test, $RF\!F$) to classify galaxies into $1$ or $2$ components are compared and investigated. As each method has different bias in selecting $2$-component galaxies, for this work we classify a galaxy as a $2$-component object if at least two methods agree in that classification, and as a $1$-component galaxy otherwise. The peculiar population is separated by visual inspection, as done in \cite{Mortlock15}. In Figure \ref{fig.sample} we show our sample of galaxies in the stellar mass and redshift space. $1$-component galaxies are further classified according to their S\'ersic index, as disc-like galaxies ($n<2.5$) or spheroid-like galaxies ($n>2.5$).

\section{Method}\label{sec.method}

\subsection{Structural parameters}\label{subsec.fitting}

In \cite{Margalef16}, we obtained the structural parameters of each galaxy in our sample from the Hubble WFC3 $H$-band images using \textsc{galfit} and \textsc{galapagos}. \textsc{galfit} is a two-dimensional fitting code used to model the surface-brightness of an object with predefined functions. The most used and useful functions to describe galaxy surface brightness profiles are the exponential profile and the S\'ersic profile \citep{Sersic} for respectively disc and bulges/spheroids systems. \textsc{galapagos} \citep{Barden12} is a software that uses \textsc{se}xtractor \citep{Bertin96} to detect and extract sources. It estimates the the sky background of each object and performs an automated S\'ersic profile fit using \textsc{galfit}.

We fit the surface brightness with two models. Model $1$ consist of a single S\'ersic profile, with $n$ as a free parameter, and represents well a $1$-component galaxy. Model $2$ describes a galaxy with two components, represented by an inner S\'ersic and an outer exponential profile. From the best fit to model $1$  systems we obtain the following parameters: position of the galaxy within the image stamp $(x,y)$, effective radius $R_e$, S\'ersic index $n$, $A\!B$-magnitude $m$, axis ratio $q$ and position angle $P\!A$, and for model $2$: position in the stamp $(x,y)$, effective radius of bulge and disc component $(R_{e\,\mathrm{B}},R_{e\,\mathrm{D}})$, S\'ersic index of the bulge $n_\mathrm{B}$, $AB$-magnitude for bulge and disc $(m_\mathrm{B},m_\mathrm{D})$, axis ratio of bulge and disc $(q_\mathrm{B},q_\mathrm{D})$ and position angle of both components $(P\!A_\mathrm{B},P\!A_\mathrm{D})$. We exclude any fitting with unphysical parameters: effective radius smaller than $0.5$ pixels, or larger than the size of the image stamp, $q<0.1$, and $n<0.5$ or $n>8$, which implies that about $8$ per cent of the galaxies are not well represented by either model $1$ or model $2$. These galaxies which are not fit are either very compact objects, or considerably faint/small, and have an average redshift of $z=2$.

\subsection{Multiwavelengh Analysis}\label{subsec.multi}

To extend our study to a broader range in wavelength than previously done, we use all of the other bands available from \textit{HST} in the CANDELS/UDS field ($J$, $i$, $V$ bands).   The reason for this is that we get a full spectral energy distribution for both the inner and outer components by just using the HST data.  Within the redshift range of $1<z<3$ the observed $H$-band samples the underlying stellar mass better than bluer wavelengths.  Within this band we obtain as close as possible a measure of the underlying stellar mass distribution.  We therefore adopt the structural parameters measured in terms of the  S\'ersic index and radius obtained from fitting to the $H$-band, which we then apply to the other wavelengths as a prior in our fitting.  This ensures that we are measuring consistent fluxes within the same physical area of the galaxy. 

There are several reasons why this approach is successful.    We note from previous studies that above wavelengths of $3000$\AA\, the structure and sizes of galaxies do not change significantly \citep{Lanyon12,Conselice11,Papovich05}, and therefore we can assume that there are no significant colour gradients. Using \textsc{galfit} we fit the surface brightness with models $1$ and $2$, with all the parameters in each model fixed to those of the best fit determined by the H-band fit, but allowing the magnitudes to vary as a free parameter. This implies for model $1$ keeping fixed the position on the stamp, the effective radius, S\'ersic index, axial ratio and position angle. For model $2$ we fix the position on the stamps, the effective radius of the disc and the bulge, the S\'ersic index of the bulge, the axial ratio and position angle for bulge and disc. The sky level for each object, in the different bands, is calculated using \textsc{galapagos}. We run \textsc{galapagos} in each band using the detections determined from the $H$-band, and obtain the local sky around all the objects in our sample, as explained in \cite{Barden12}.

\subsection{SEDs}\label{subsec.SEDs}

Once we obtain a full SED for our components, the stellar masses are measured using SMpy \citep{Duncan14}, a template fitting code with SEDs derived from the synthetic stellar population models of \cite{Bruzual03}. The model SEDs are generated from a single stellar population. We use the initial mass function (IMF) of \cite{Chabrier03}, and we use the extinction law of \cite{Calzetti00} to include dust in the templates. The star formation history (SFH) is characterised by an exponentially declining model with various ages, metallicities and dust extinctions

\begin{equation}
 SF\!R(t) \propto SF\!R_{0}\, e^{-t/\tau},
\end{equation}

\noindent where the values of $\tau$ ranges between $-10$ to $10$ Gyr (negative $\tau$ values represent exponentially increasing histories). Age is allowed to vary from $0.02$ to $9.6$ Gyr, dust ranges from $0$ to $4$, and metallicities from $0.0001$ to $0.05$.

As we show in Appendix \ref{appendix}, to be able to obtain reliable rest-frame colours, SFRs and stellar masses from the SED fitting, we need  photometry not only in the four \textit{HST} bands but also in the $U$ band, $K$ band and Spitzer channel $1$ at $3.6\rm{nm}$ ($ch1$). However, we do not have high enough resolution images in these bands to perform our bulge to disc decompositions for our sample of galaxies. 

We therefore make use of the photometric catalogue from \cite{Galametz13}. The photometry from this catalogue is performed in a different way than we do, and uses different apertures than we do in this work, and so, we cannot simply combine both catalogues. However, as there are no significant colour gradients for these galaxies, it is a fair assumption that colours (in particular $U-V$, $H-K$, $H-ch1$) are the same in both catalogues, that is:
\begin{equation}
(U_{this\ work}-V_{this\ work})=(U_{Galametz13}-V_{Galametz13})
\end{equation}
\begin{equation}
(H_{this\ work}-K_{this\ work})=(H_{Galametz13}-K_{Galametz13})
\end{equation}
\begin{equation}
(H_{this\ work}-ch1_{this\ work})=(H_{Galametz13}-ch1_{Galametz13})
\end{equation}

\noindent We then obtain the magnitudes we would derive if we have measured resolved imaging on the missing bands -- i.e.\  $U_{this\ work}$, $K_{this\ work}$ and $ch1_{this\ work}$, and incorporate them to our photometric catalogue. In the Appendix \ref{appendix} we explore how our results from the SED fitting improve by adding these three bands to our photometry.

For the $2$-component systems, we make another assumption, due to the fact that the \cite{Galametz13} catalogue does not have bulge to disc decomposed photometry. We first estimate, from a sample of nearby galaxies (see Appendix \ref{appendix} for details), how the B/T changes as a function of wavelength over a broad range, and use this information along with the previously calculated magnitudes ($U_{this\ work}$, $K_{this\ work}$ and $ch1_{this\ work}$) to infer the magnitudes for the bulge and disc component in $K$, $ch1$ and $U$ band.

\begin{figure}
  \includegraphics[width=1\linewidth]{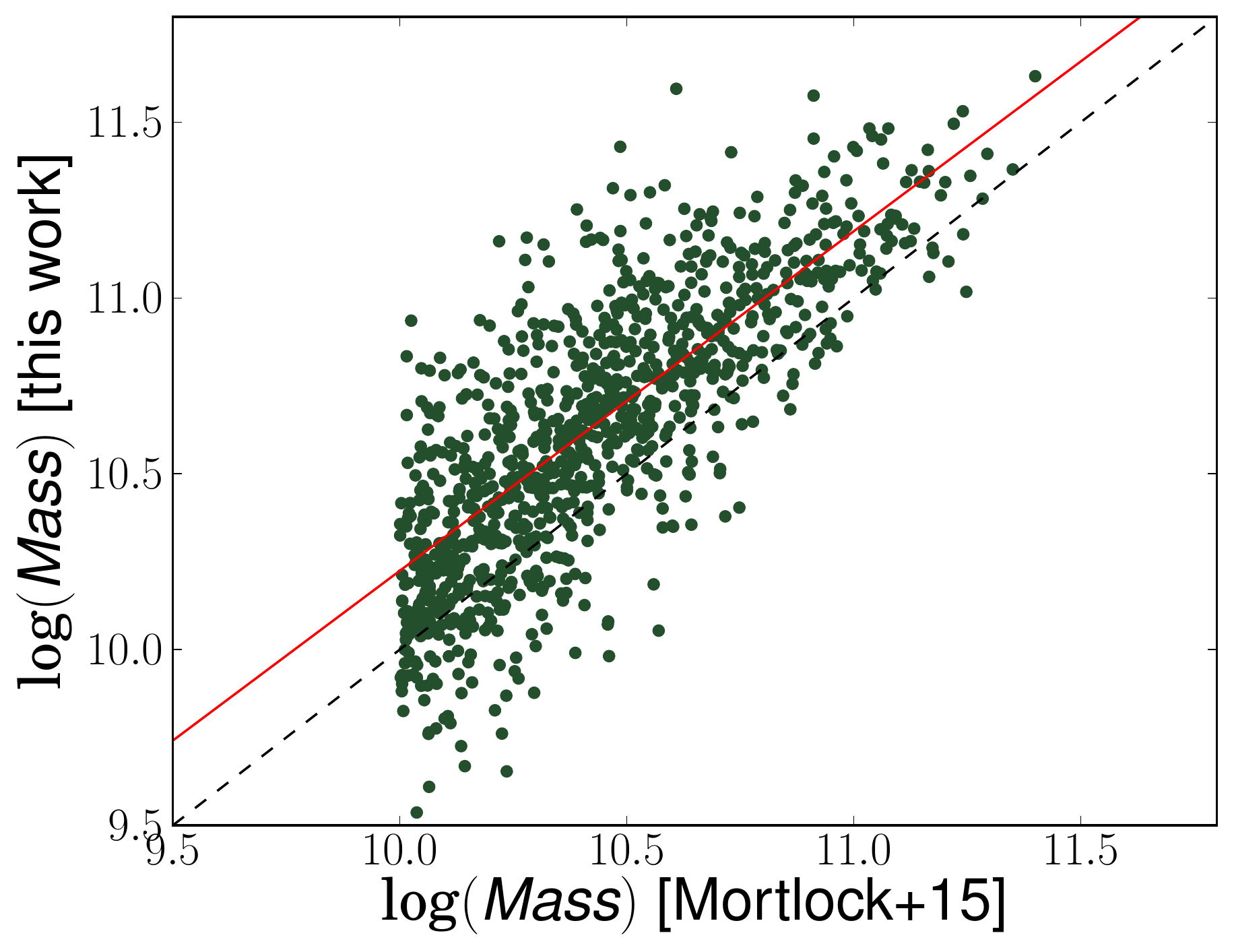}
  \caption{Comparison between the stellar masses calculated in this work and in \protect\cite{Mortlock15}. The galaxies in this plot have their masses calculated assuming that the light profiles are well fit by a single S\'ersic profile. The dashed line is a 1:1 relation between the two measures of the stellar masses. The red line is the best fit line to the data.}\label{fig.mass_alice_vs_mass}
\end{figure}

In Figure \ref{fig.mass_alice_vs_mass} we compare the stellar mass from the initial catalogue we selected our sample from \cite{Mortlock15} to those masses calculated in this work, assuming all galaxies have a single component. On average the stellar masses calculated in this work are $0.2$dex larger. This systematic offset found between the two mass estimates could arise from using different fitting codes and free parameters. The large scatter may be cause by the errors in the stellar masses due to degeneracies in the fitting code. Through the rest of the paper we use these new masses.

\begin{figure*}
  \includegraphics[width=0.49\linewidth]{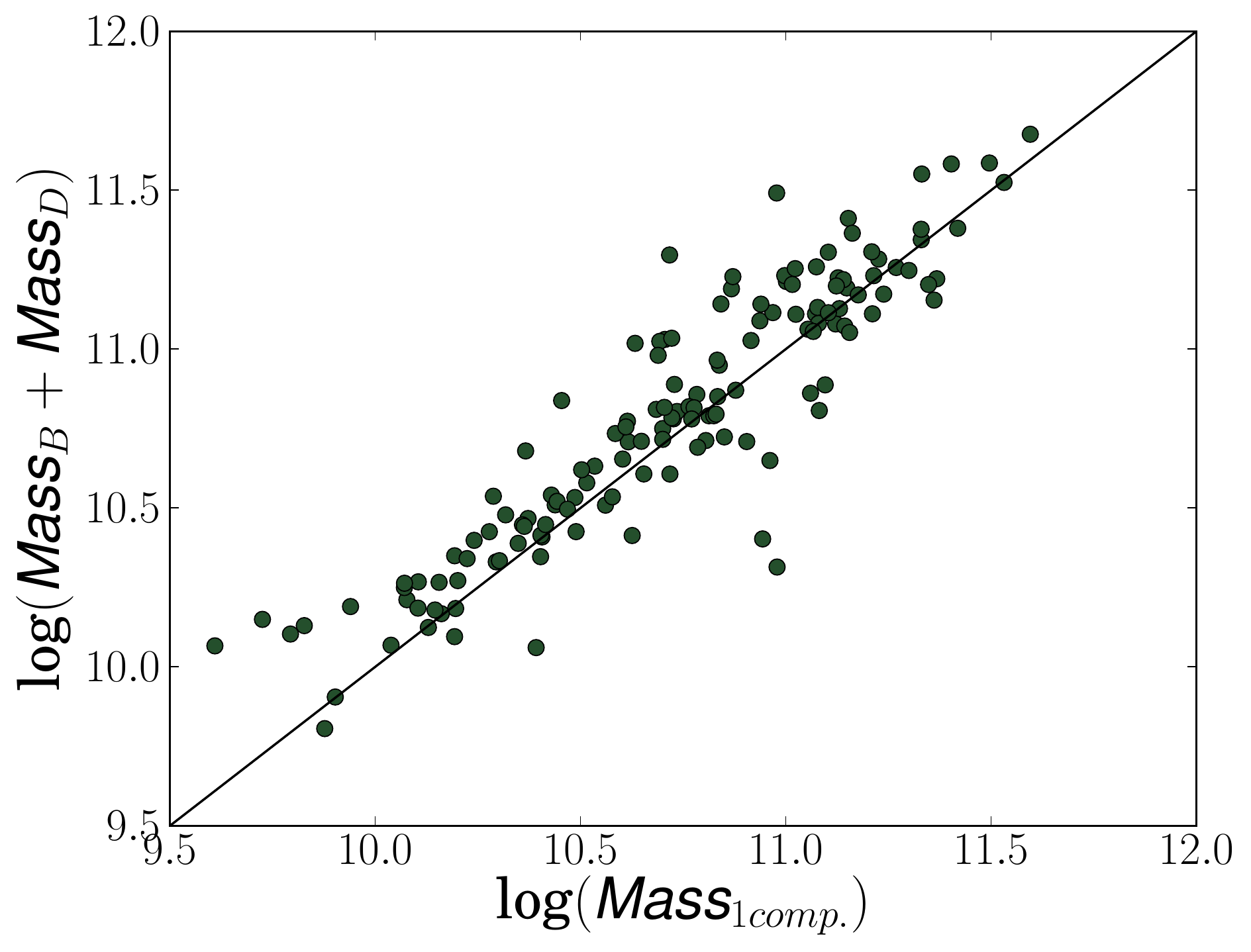}
  \includegraphics[width=0.49\linewidth]{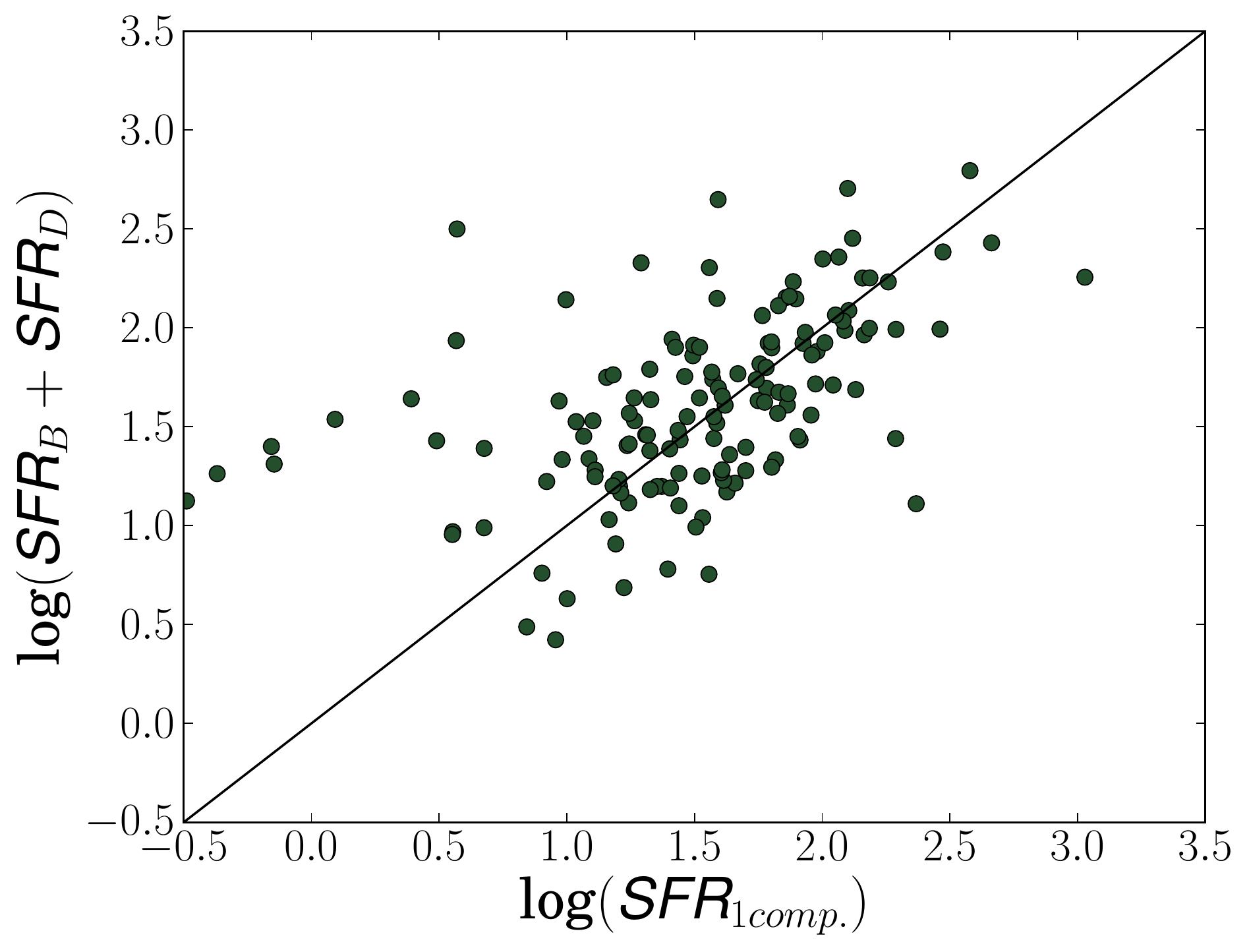}
  \caption{Comparison between the $2$ and $1$ component models for galaxies classified as $2$-~component systems. Left: Total stellar mass of the galaxies as given by the mass of the bulge and disc ($Mass_B+Mass_D$) from the $2$-component model compared to the total mass of the galaxy obtain from the $1$-component model. Right: total $SF\!R=SF\!R_B+SF\!R_D$ from the $2$-component model compared to the $SF\!R$ obtained from the $1$-component fitting.}\label{fig.2c_vs_1c}
\end{figure*}

For the $2$-component galaxies we have photometry for both the bulge and the disc, and therefore we are able to perform SED fitting for each component separately, allowing us to measure stellar masses and SFRs within both components. In Figure \ref{fig.2c_vs_1c} we compare the stellar mass (left) and SFR (right) of the $2$-component galaxies we measure, as if their light is in a single component, and as the sum of the stellar mass/SFR of each separate components. We find a good correlation between the stellar masses measured between the $1$- and $2$-component systems. There is also a good correlation between the star formation measured, but with a higher scatter.

\subsection{Star Formation Rates}\label{subsec.method.SFR}

The rest-frame $UV$ is a good SFR indicator, as it traces the presence of young and short lived O and B stars produced by recent star formation. Furthermore, the $UV$ is the only star formation indicator which we can use to resolve the SFHs in galaxies \citep[e.g.,][]{Ownsworth12}. By applying a scaling factor of the $UV$ luminosity, the SFR can be calculated, however, this depends on the assumed IMF  \citep[e.g.,][]{Kennicutt83}. The downside of calculating the SFR from the $UV$ luminosity is its susceptibility to dust extinction, and therefore a dust correction has to be applied. We use a correction based on the $UV$ slope, which is explained below.

To calculate the SFR we obtain the rest-frame $UV$ luminosities (at $2800\angstrom$) from the best fit SED model of the stellar mass fitting, we then use the \cite{Kennicutt98} conversion from $2800\angstrom$ luminosity to SFR assuming a Chabier IMF:

\begin{equation}
 SF\!R_{UV}(\mathrm{M}_{\odot}\mathrm{yr}^{-1})=8.24 \cdot 10^{-29}L_{2800}(\mathrm{ergs}\ \mathrm{s}^{-1}\mathrm{Hz}^{-1}).
\end{equation}

\noindent This gives the SFR before a dust correction is applied.

\subsubsection{Dust Correction}\label{subsubsec.dust}

We use the $UV$ slope as an estimate for the dust extinction. \cite{Meurer99} found a correlation between the $UV$ slope $\beta$ and the attenuation due to dust. Following the method described in \cite{Ownsworth16} we use ten $UV$ windows defined by \cite{calzetti94} to measure the $\beta$ by interpolating the flux in these windows from the SED fitting. We then convert the slope into a dust correction using the \cite{Fischera05} dust model
\begin{equation}
A_{2800}=1.67\beta+3.71.
\end{equation}

However, the $UV$ slope can also be affected by the age of the stellar population. Galaxies with old and passive stellar populations may have similar SEDs as highly obscured galaxies with young star-forming populations. This means that passive galaxies can appear to be highly dust obscured, and thus we would derive higher SFR than they actually have.

To solve this problem we treat differently the passive and star-forming galaxies as selected by the $UV\!J$ diagram (see \S \ref{subsec.colours}). We only use these dust corrections calculated by the $UV$ slope for galaxies selected as star-forming within the $UV\!J$ diagram. For the passive population we assume that star-forming and passive galaxies have similar dust masses \citep[e.g.,][]{Bourne12}. We therefore assume that on average passive galaxies have the same dust attenuation as star-forming ones within a range of redshift and similar stellar masses. 
 
\section{Results}\label{sec.results}

In the following section we discuss the properties of the inner and outer components of our galaxy sample. We first describe the colours and the location of our components on a $UV\!J$ colour-colour diagram to put some constrains on the passivity and star formation presence within discs and bulges. We later discuss the directly measured SFRs of these galaxy components, as well as the stellar masses, and how these properties change with time. 

\subsection{Colours}\label{subsec.colours}

Studying the colours of galaxies gives us information about the star formation activity of each component, as blue objects indicate star formation activity, while red colour signifies that a galaxy is passive or dusty. The $U$, $V$ and $J$ rest-frame magnitudes are calculated from the flux measured by the $U$, $V$ and $J$ Bessel filters in the best SED model, during the SED fitting. We divide our sample into star-forming and passive galaxies using the rest-frame $UV\!J$ colours, where a galaxy is classified as red/passive if it satisfies the following criteria \citep[see][]{Mortlock13},

\begin{equation}\label{eq.uvj}
\begin{cases}
 (U-V)>1.3 \\
 (V-J)<1.6 \\
 (U-V)>0.88 (V-J)+0.49
\end{cases} 
\end{equation}

\noindent and as blue otherwise.

\begin{figure}
  \includegraphics[width=1\linewidth]{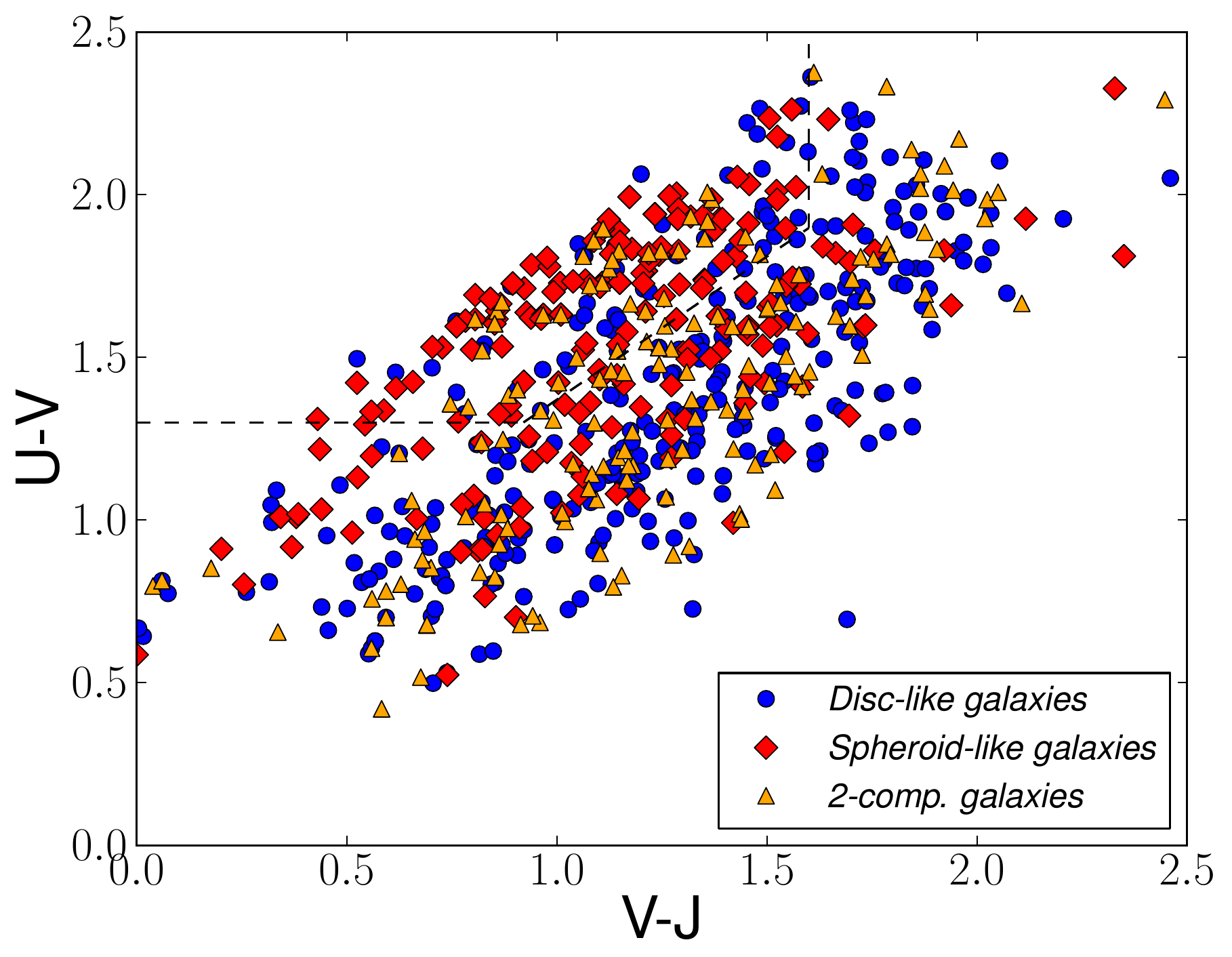}
  \caption{$UV\!J$ diagram for $1$-component galaxies. Red circles represent spheroid-like galaxies ($n>2.5$) and blue circles, disc-like ones ($n<2.5$). The box in the left top corner represents the passive population, given by equations \ref{eq.uvj}.}\label{fig.uvj1}
\end{figure}

Figure \ref{fig.uvj1} shows where the $1$-component galaxies in our sample lie in the $UV\!J$ diagram. The majority of disc-like galaxies, $78 \pm 6$ per cent, are in the star-forming region of the $UV\!J$ diagram, while $67 \pm 6$ per cent of spheroid-like galaxies are in the passive region. This is consistent with the idea that galaxy morphology correlates with colour, and therefore star formation activity, with disc-dominated galaxies being more likely to be blue and star-forming, while elliptical or spheroid-dominated galaxies are passive. However, note that this is only for galaxies which are best fit as single components, which in the nearby Universe would be early-type ellipticals and pure disc galaxies.

\begin{figure*}
  \includegraphics[width=0.49\linewidth]{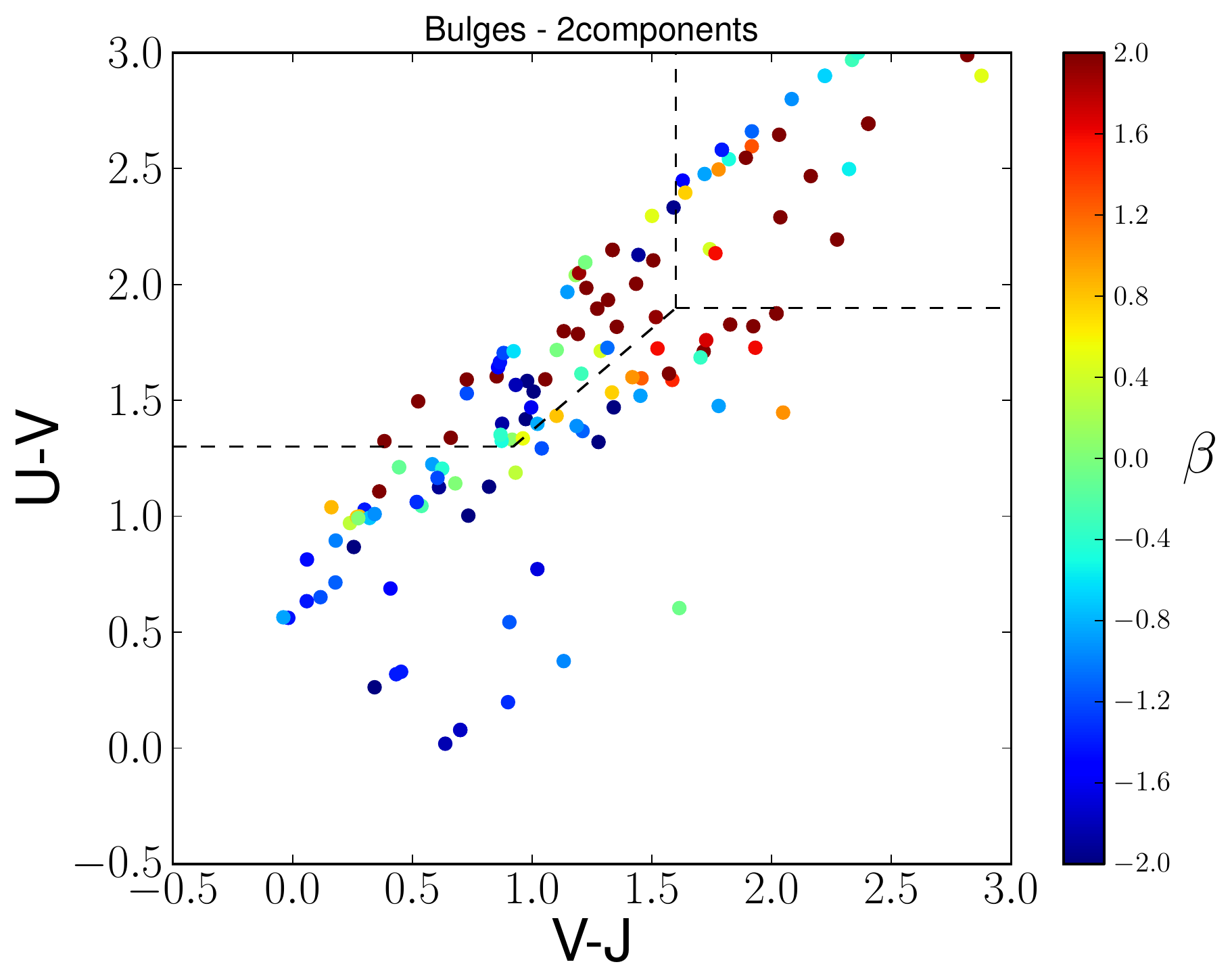}
  \includegraphics[width=0.49\linewidth]{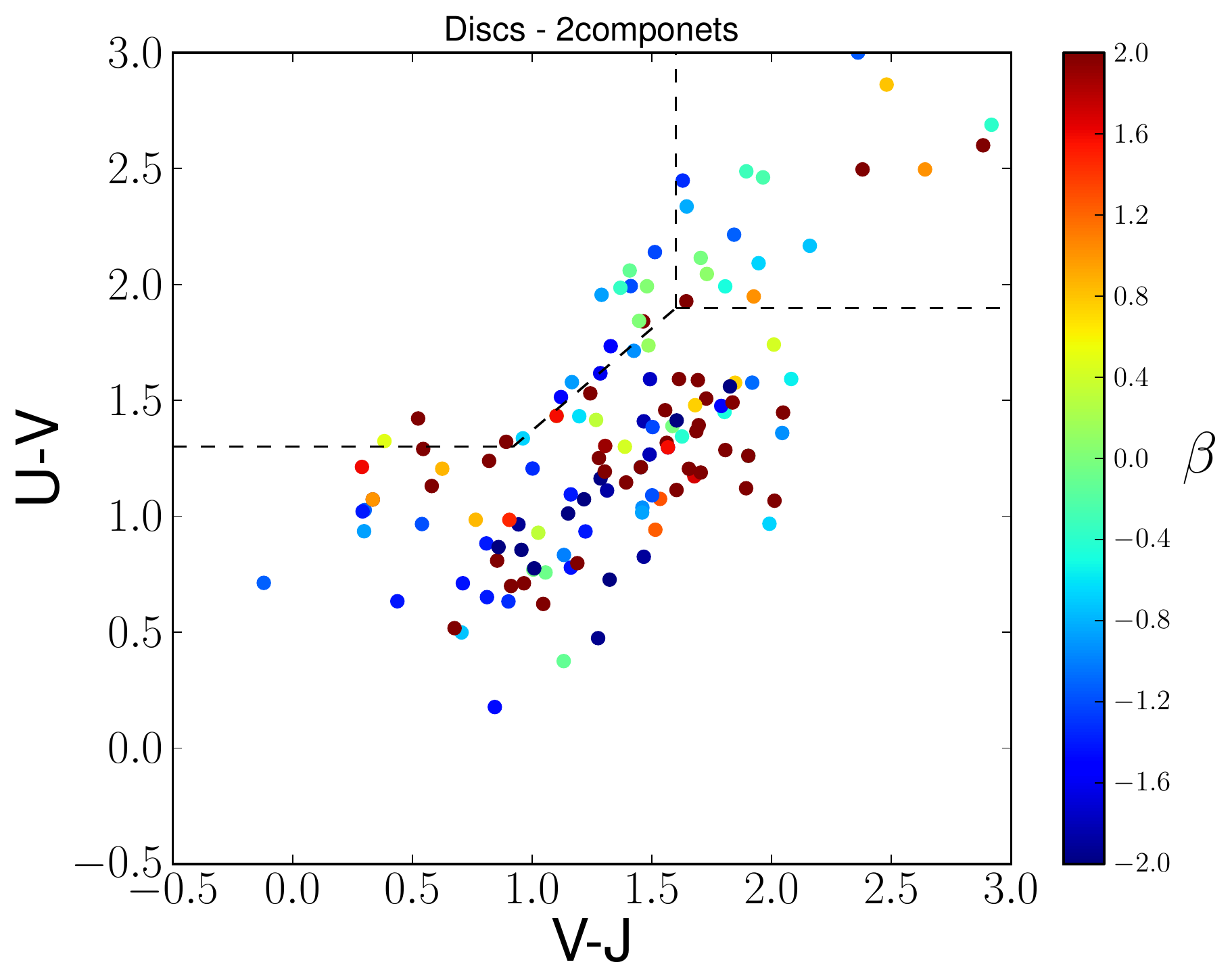}
  \caption{$UV\!J$ diagram for bulges (left) and discs (right) of the $2$-component galaxies. The box in the left top corner represents the passive population (equation \eqref{eq.uvj}), and in the right top corner we find the dusty region. The colour bar represents the $UV$ slope $\beta$ derived from our SED fitting. As can be seen, there is a difference between the stellar population of inner and outer components even out to $z \sim 2$.}\label{fig.uvj2}
\end{figure*}

We see something similar, but with some differences, when we look at the $UV\!J$ colours of the disc and bulge components. As can be seen in Figure \ref{fig.uvj2}, the majority of both bulges and discs are within the star-forming region ($90 \pm 11$ and $68 \pm 9$ per cent respectively). However, a significant fraction of star-forming bulges, $30$ per cent, are in the dusty region of the diagram, while only $17$ per cent of star-forming discs lie in that region. The inner parts of $2$-component galaxies are therefore red due to the dust according to the $UV\!J$ diagram, rather than because of an old stellar population. The fraction of passive and star-forming objects are summarize in Table \ref{table.uvj}.

\begin{table}
\centering  
	\begin{tabular}{lrr} 
                        \hline
		&Star-forming & Passive \\[0.5ex] \hline \hline
			\rule{0ex}{2.8ex}$1$-comp. discs  & $78 \pm 6\%$ & $22 \pm 3\%$  \\ \hline
			\rule{0ex}{3ex}$1$-comp. spheroids  & $43 \pm 5\%$ & $67 \pm 6\%$ \\
			\hline \hline                 
			\rule{0ex}{3ex}$2$-comp. discs  &  $90 \pm 11\%$ & $10 \pm 3\%$  \\ \hline
			\rule{0ex}{3ex}$2$-comp. bulges  &  $68 \pm 9\%$ & $32 \pm 5\%$ \\       \hline
	\end{tabular} 
	\caption{Fraction of galaxy components that are passive and star-forming, according to the $UV\!J$ selection given by equations \eqref{eq.uvj}.}\label{table.uvj} 
\end{table}

\begin{figure*}
  \includegraphics[width=0.49\linewidth]{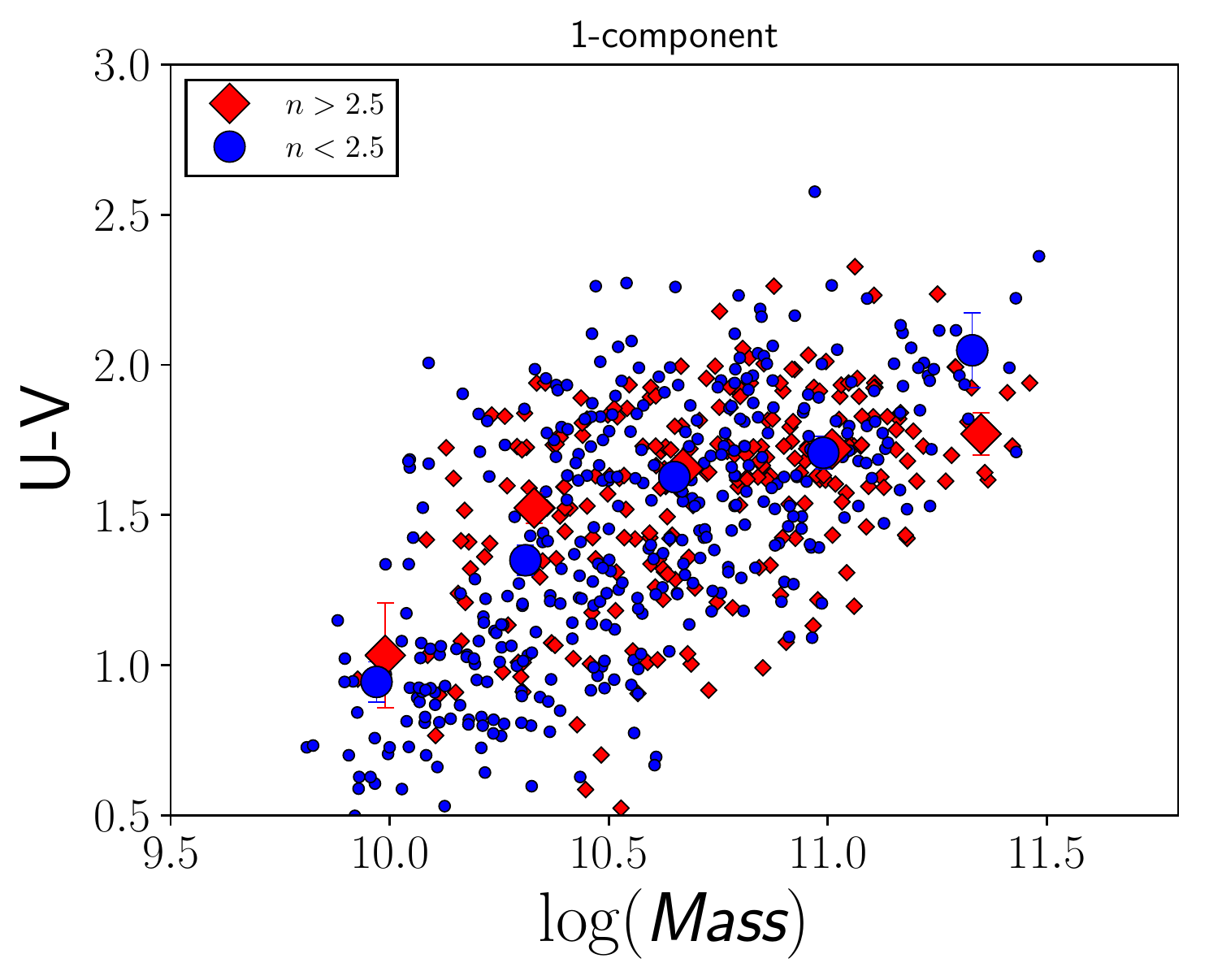}
  \includegraphics[width=0.49\linewidth]{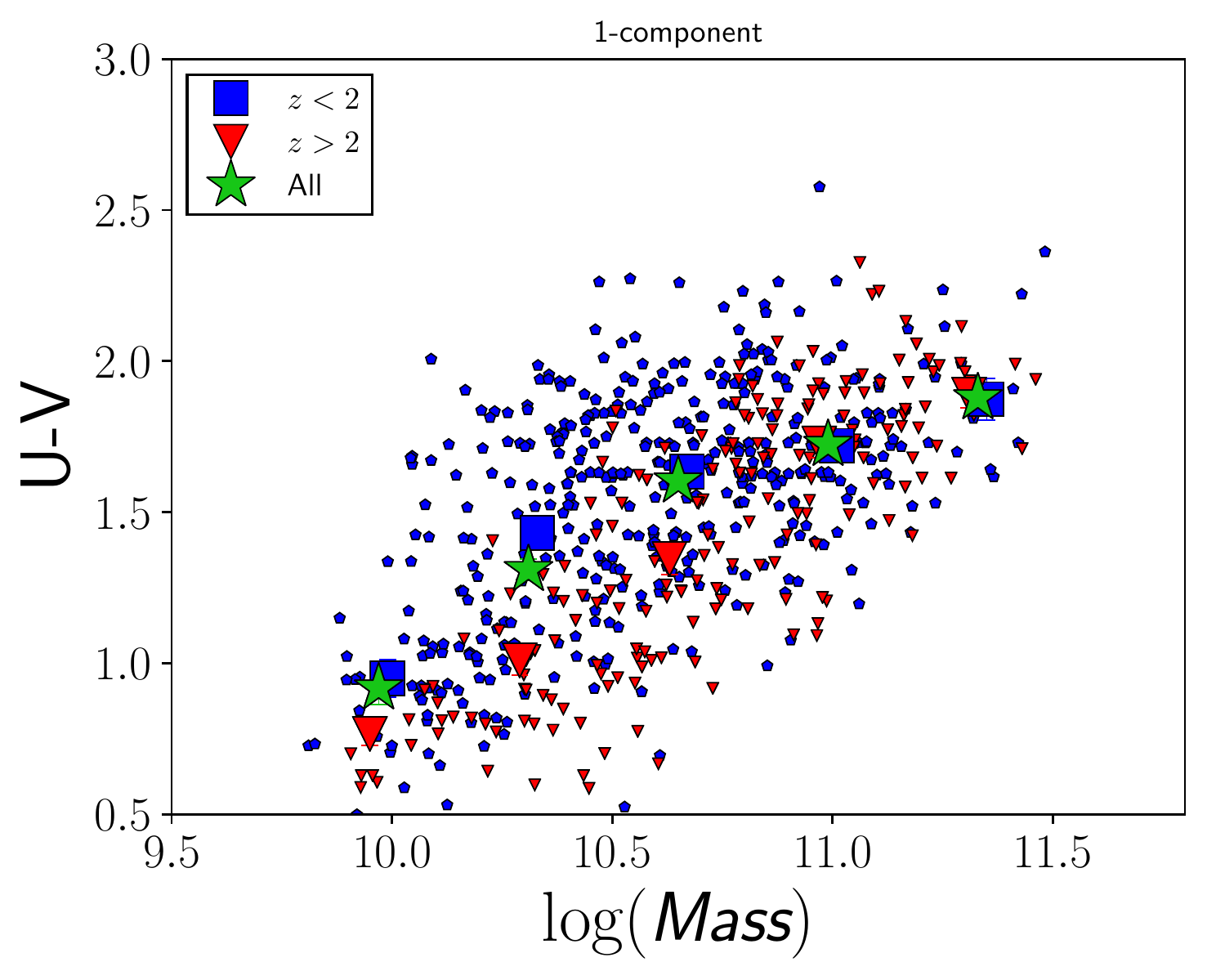}\\
  \includegraphics[width=0.49\linewidth]{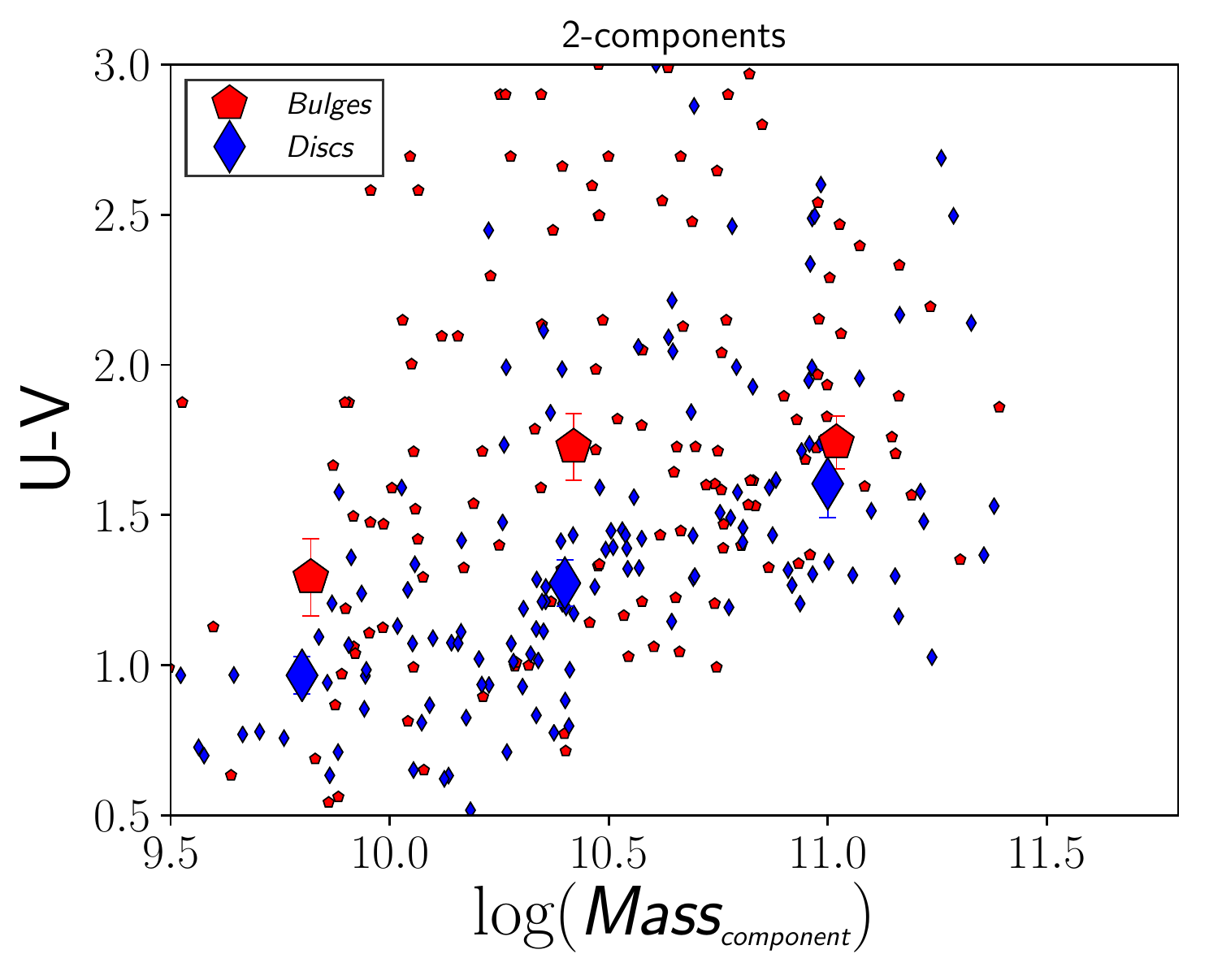}
  \includegraphics[width=0.49\linewidth]{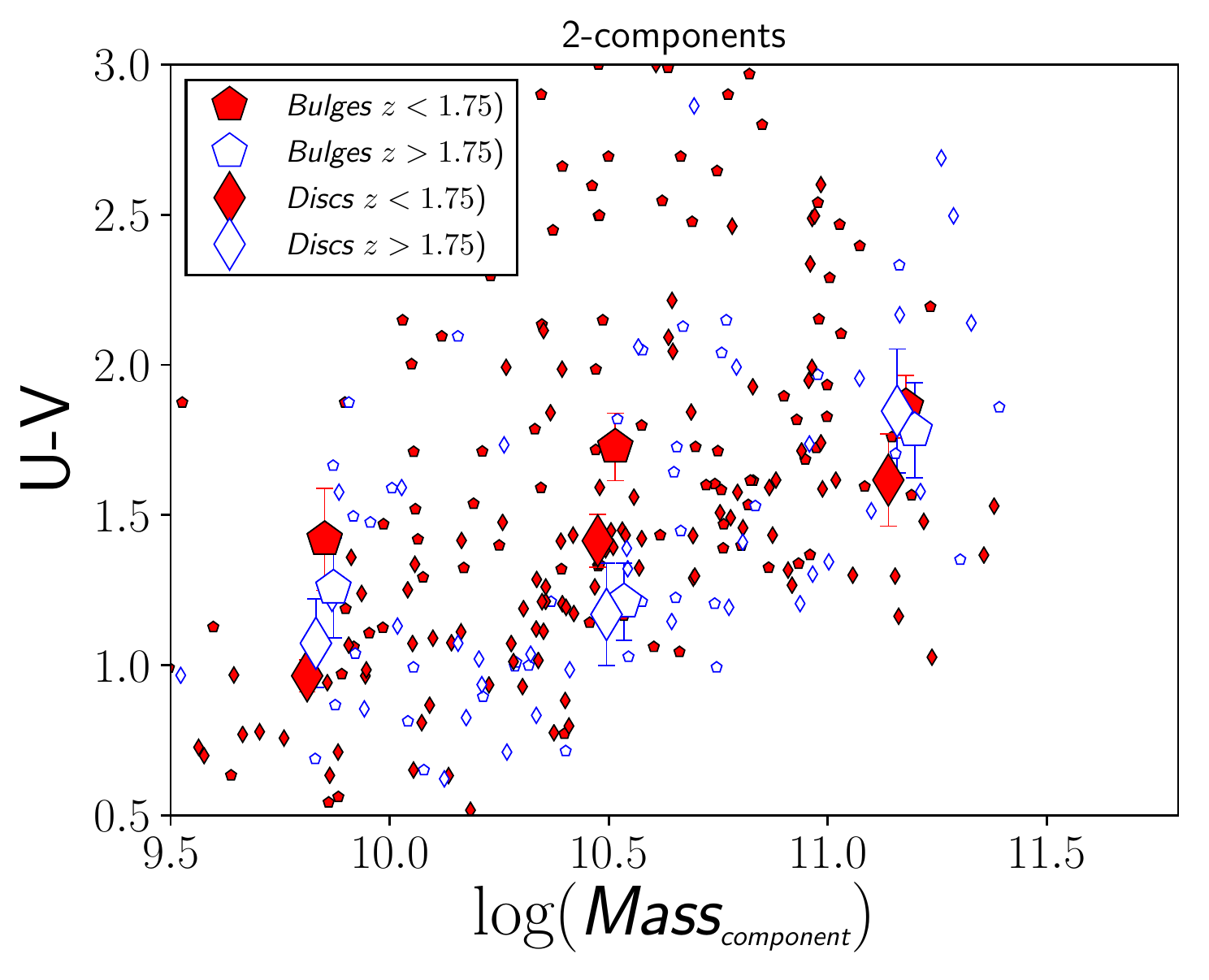}
  \caption{Rest-frame colour $U-V$ as a function of stellar mass. On the top panels we show the $1$-component galaxies separated by S\'ersic index (left), where blue diamonds are galaxies that have $n>2.5$ (i.e.\ spheroid-like) and blue circles are for galaxies with $n<2.5$ (i.e.\ disc-like); and by redshift (right), blue squares are galaxies at $z<2$, and red triangles are galaxies at $z>2$. On the bottom panels we show the $2$-component galaxies. Left: $U-V$ colour of bulges and discs as a function of the stellar mass of the component. Right: bulges and discs of $2$-component galaxies separated in two redshift bins. The big symbols represent the median values.}\label{fig.cuv_vs_mass}
\end{figure*}

We furthermore study the rest-frame $U-V$ colour as a function of stellar mass and redshift (Figure \ref{fig.cuv_vs_mass}). We find that there is a trend with stellar mass for all types of galaxies and components, such that the rest-frame $U-V$ colour becomes redder at higher stellar masses, as seen in Figure \ref{fig.cuv_vs_mass}, where we plot the rest-frame $U-V$ colour as a function of stellar mass for the $1$-component galaxies (top panels) and for the $2$-component galaxies (bottom panels). On the left side we show the effect of the S\'ersic morphology, and on the right we investigate the relation between the $U-V$ colour and stellar mass as a function of redshift. 

\begin{figure*}
  \includegraphics[width=0.49\linewidth]{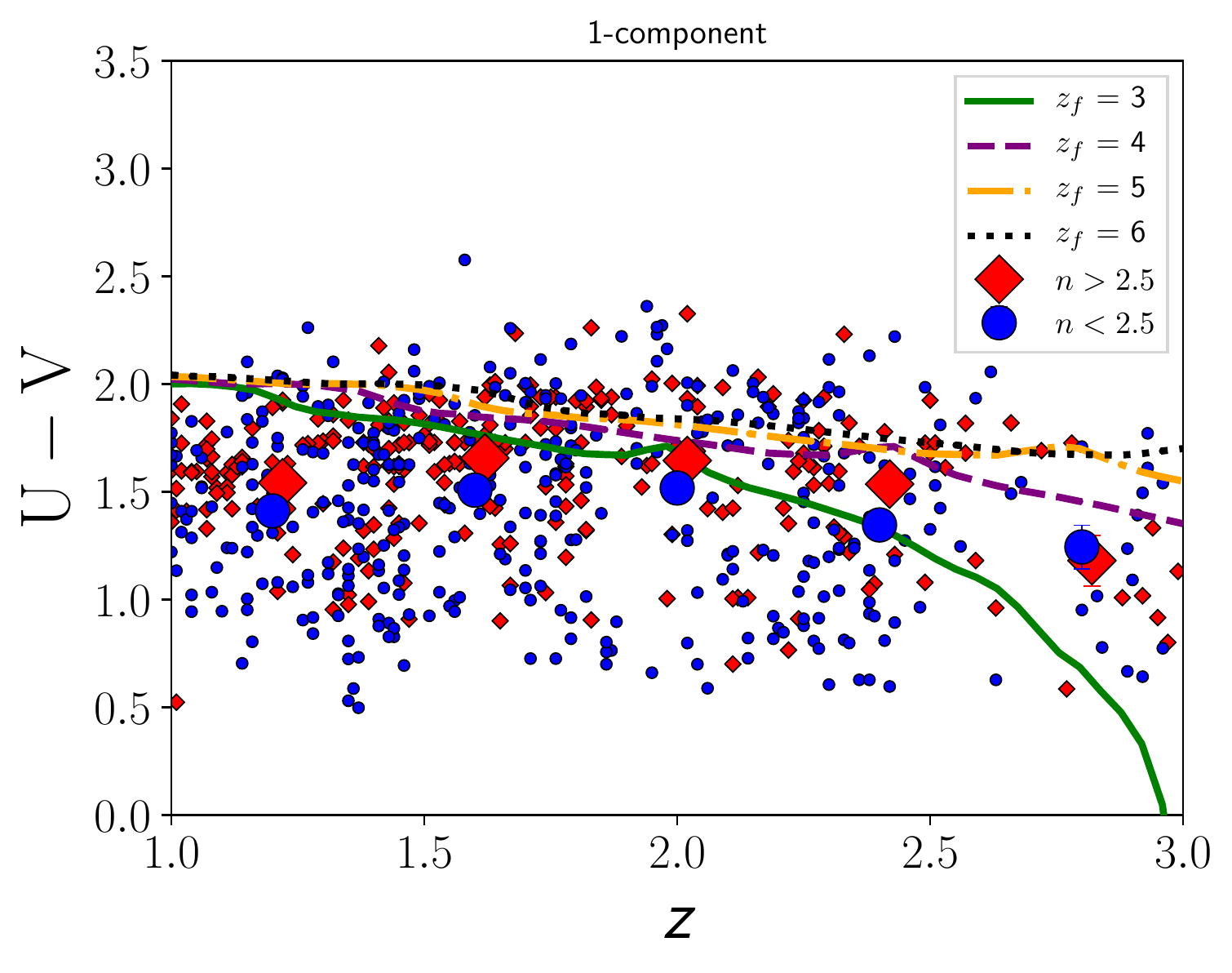}
  \includegraphics[width=0.49\linewidth]{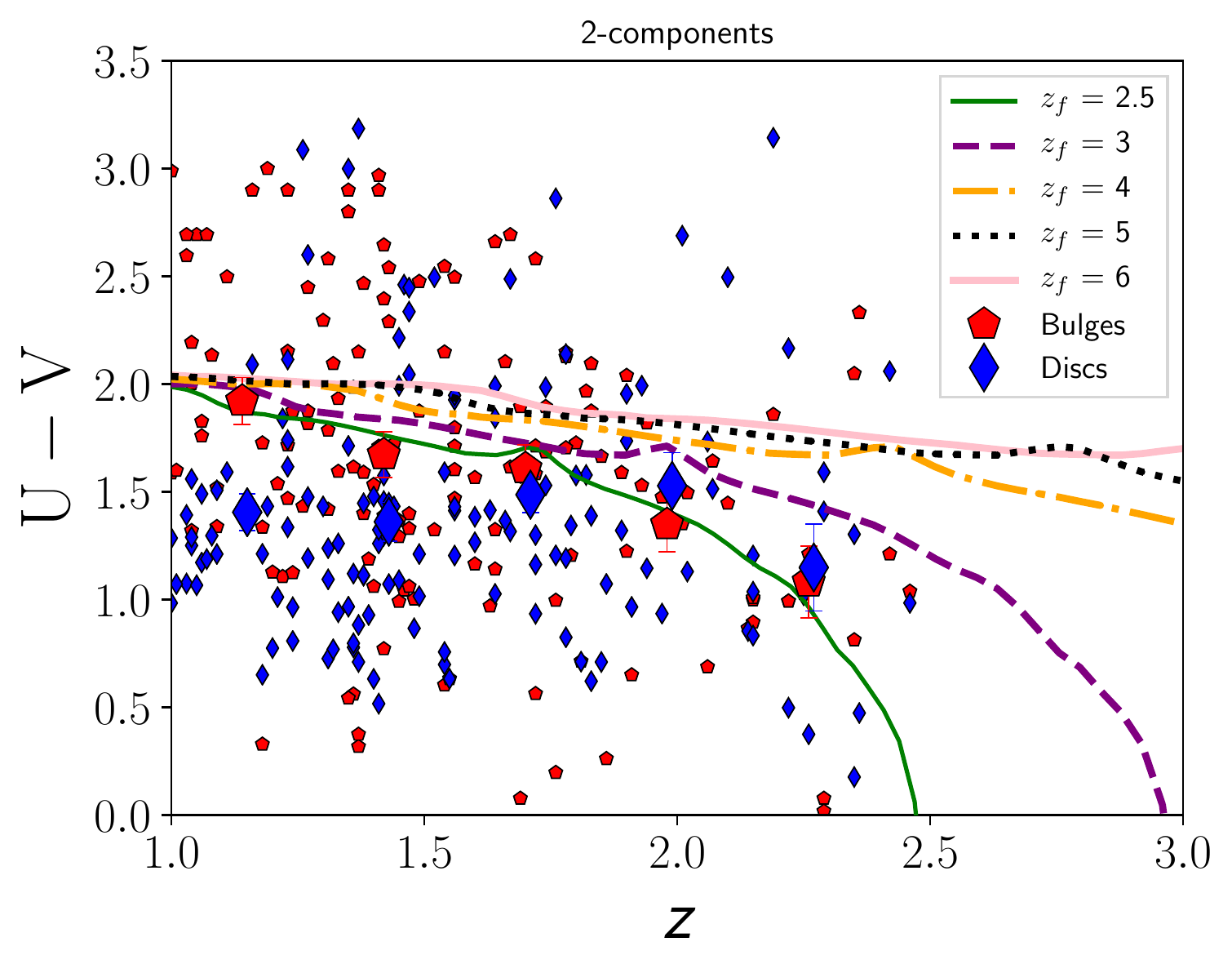}
  \caption{Rest-frame colour $U-V$ as a function of redshift. Left panel: $1$-component galaxies separated by S\'ersic index, where blue circles are disc-like galaxies ($n<2.5$), and red diamonds are spheroid-like ones ($n>2.5$). Right panel: $2$-component galaxies, with bulges represented as red pentagons and discs as blue diamonds.}\label{fig.cuv_vs_z}
\end{figure*}

For the $1$-component galaxies, the structure of a galaxy, as measured through the fitted S\'ersic index, has a small effect on the colour  over all redshifts. That is, we find the $U-V$ colour to be similar between disc-like and spheroid-like systems, but there are some differences. In Figure \ref{fig.cuv_vs_z} (left panel) it appears that spheroid-like single component galaxies have slightly redder colours that disc-like galaxies at all redshifts, however this is the case only for galaxies with $\log M_{\ast}<11$ (as seen in Figure \ref{fig.cuv_vs_mass}). This apparent discrepancy arise from the fact that in Figure \ref{fig.cuv_vs_z} the colours of galaxies with different masses are averaged, and galaxies with $\log M_{\ast}<11$ are more numerous that the most massive ones. 

Nonetheless, the colour of disc-like and spheroid-like galaxies is comparable, and therefore, $1$-component galaxies must have similar SFHs at similar stellar masses (Figure \ref{fig.cuv_vs_z}). We also observe that the colour changes with redshift significantly, particularly for galaxies with $\log M_{\ast}<11$; lower mass galaxies become redder with time. The most massive galaxies are also the reddest and do not seem to become bluer with decreasing redshift. This shows that the most massive galaxies establish their colour earlier than lower mass galaxies. 

For the $2$-component galaxies we also observe a difference in the evolution of colours according to the stellar mass of the components. The most massive bulges and discs ($\log M_{\ast}>11$) have similar colours which do not change with redshift (Figure \ref{fig.cuv_vs_mass}). Less massive bulges and discs  ($\log M_{\ast}<11$) change in colour in different ways through cosmic time. Although they may have a similar initial formation, as seen by their comparable colours at $z>1.75$, at lower redshifts ($z<1.75$), bulges become redder than discs. Therefore, low redshift bulges are redder than high redshift bulges, showing little new star formation. There is however no significant evolution of disc colour with redshift, for a given mass, which implies a continuous star formation as seen in Figure \ref{fig.cuv_vs_z}, where we also observe how bulges become redder and therefore more passive.

In Figure \ref{fig.cuv_vs_z} we overplot the $U-V$ colour evolution tracks derived from \cite{Bruzual03} single stellar population models formed with a burst in star formation at different redshifts ($z=2.5$, $3$, $4,5$, $6$). The average rest-frame $U-V$ colour for bulges of $2$-component galaxies is consistent with a galaxy that had a single star formation burst at $z=2.5$. The discs of $2$-component galaxies have bluer colours than any of the model tracks which implies that they must have undergone continuous star formation activity, and the same is seen in the $1$-component galaxies.

\subsection{Star Formation Rates of Inner and Outer Components}\label{subsec.sfr}

\begin{figure*}
  \includegraphics[width=0.49\linewidth]{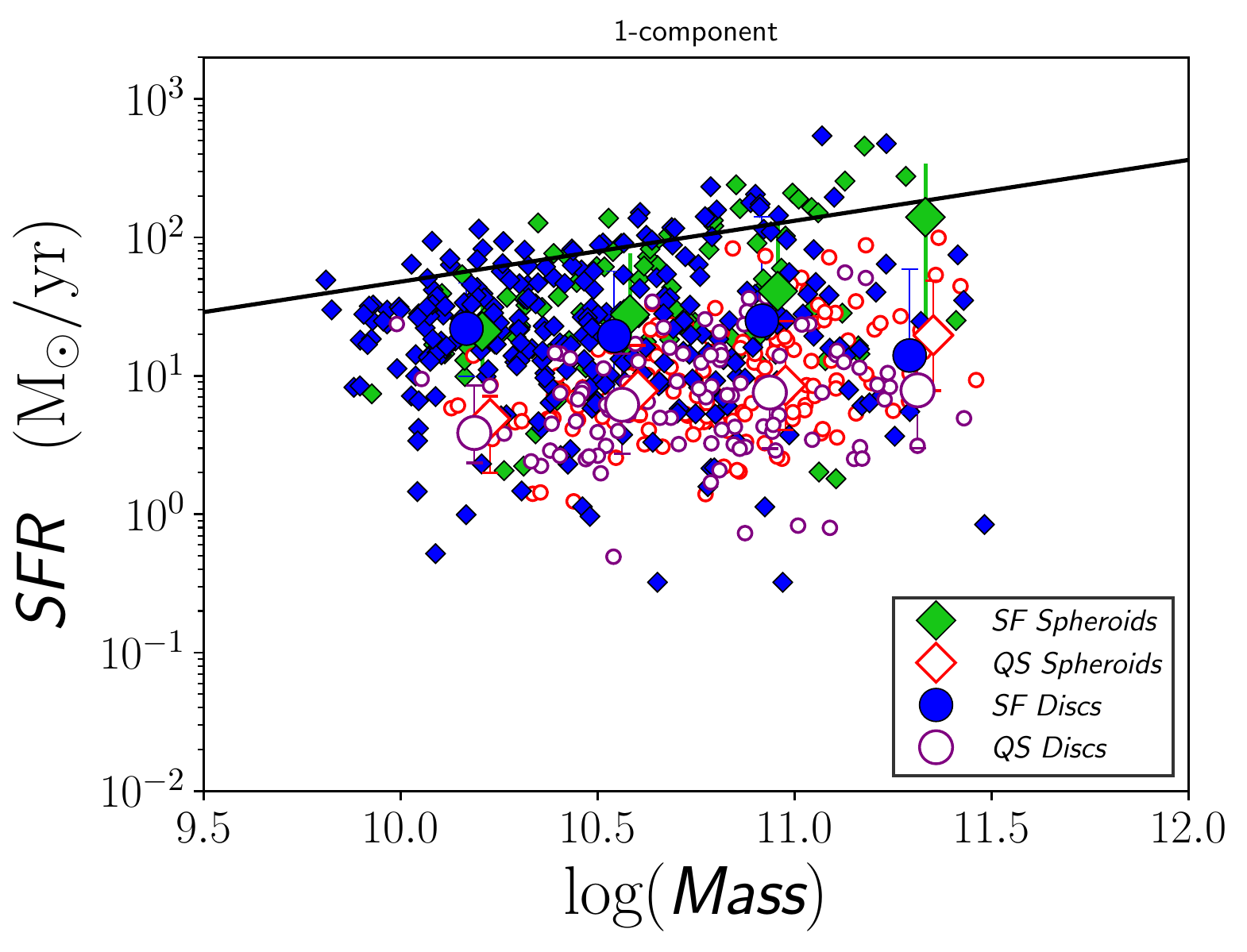}
  \includegraphics[width=0.49\linewidth]{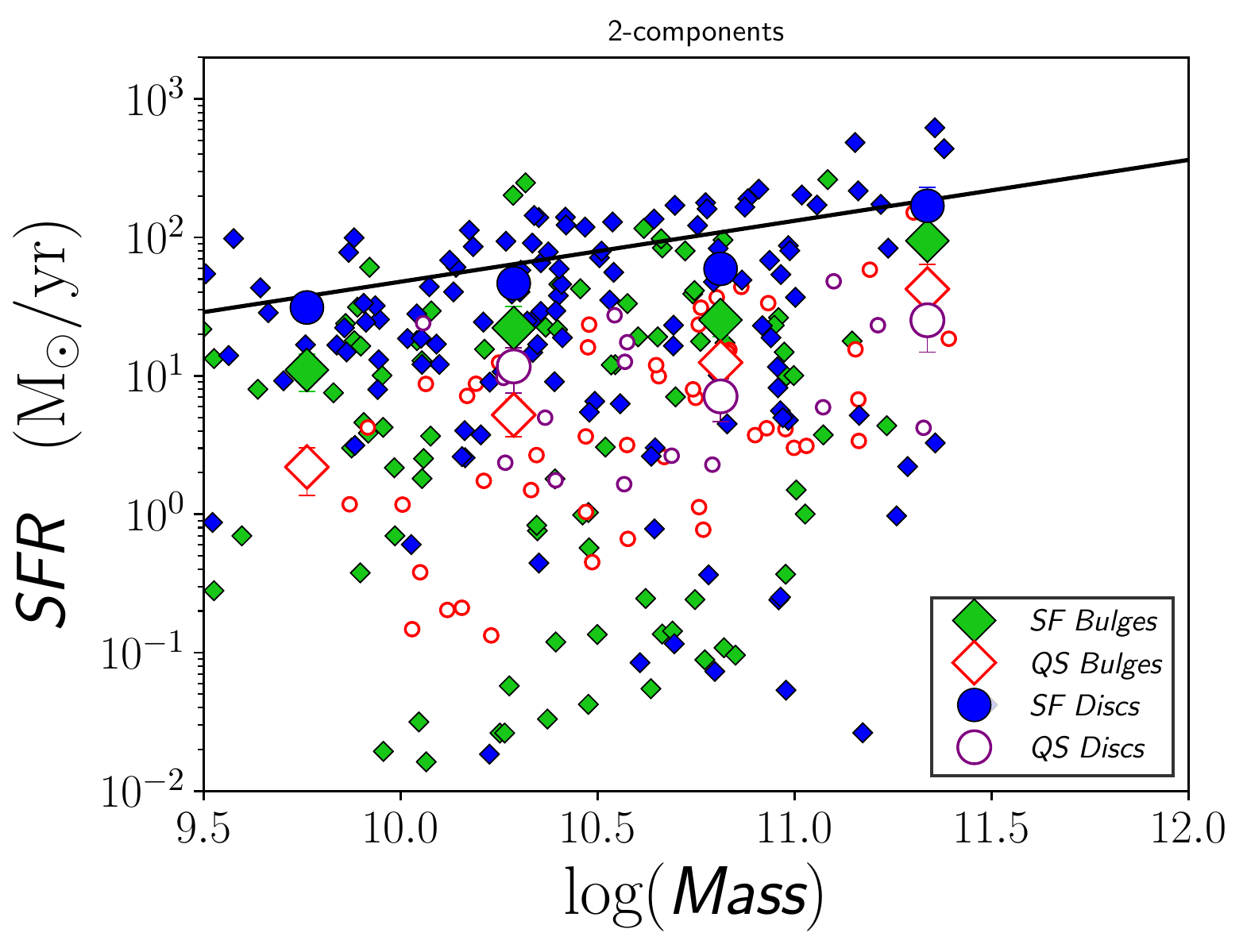}
  \caption{SFR as a function of stellar mass for $1$-component galaxies (left) and $2$-component systems (right), separated by their position on the $UV\!J$ diagram (i.e.\ passive or star-forming). The black lines are the SFR-mass sequence for star-forming galaxies found by \protect\cite{Whitaker12} at $z=2$. We show the average values with the big symbols as a representative value.}\label{fig.SFR_vs_mass}
\end{figure*}

From previous works we know that there is a tight relation between stellar mass and SFR for star-forming galaxies, which is sometimes called the `main-sequence' \citep{Noeske07}, with SFR increasing with stellar mass.  In our systems we find that $1$-component galaxies have SFRs which increases with stellar mass (Figure \ref{fig.SFR_vs_mass}). In Figure \ref{fig.SFR_vs_mass}  we show that a star-forming `main sequence'-like correlation exists for disc-like  and spheroid-like galaxies. For $2$-component galaxies, the star-forming discs have an average SFR of $SF\!R=54\ M_{\ast}\mathrm{yr^{-1}}$, in agreement with the main sequence found by \cite{Whitaker12}. The star-forming bulges have an average of $SF\!R=20 \ M_{\ast}\mathrm{yr^{-1}}$, and contain a larger dispersion, with a value of the standard deviation normalised by the mean of $\sigma_{SFR}/\langle SFR\rangle=2.1$, while the star-forming discs the ratio is $\sigma_{SFR}/\langle SFR\rangle=1.3$.  This shows that whilst the inner components on average have a trend of an increased SFR with stellar mass, the dispersion is much larger than for the outer or disc components. This likely reflects the variety of star formation and formation histories for inner or bulge components in galaxies.

This implies that the production of the main-sequence is driven by the formation of discs, and that the observed correlation between stellar mass and star formation for $2$-component systems is driven by the formation of their outer portions, or discs. This is likely because the discs contain most of the gas which is converted to stars through star formation.

\begin{figure}
  \includegraphics[width=1\linewidth]{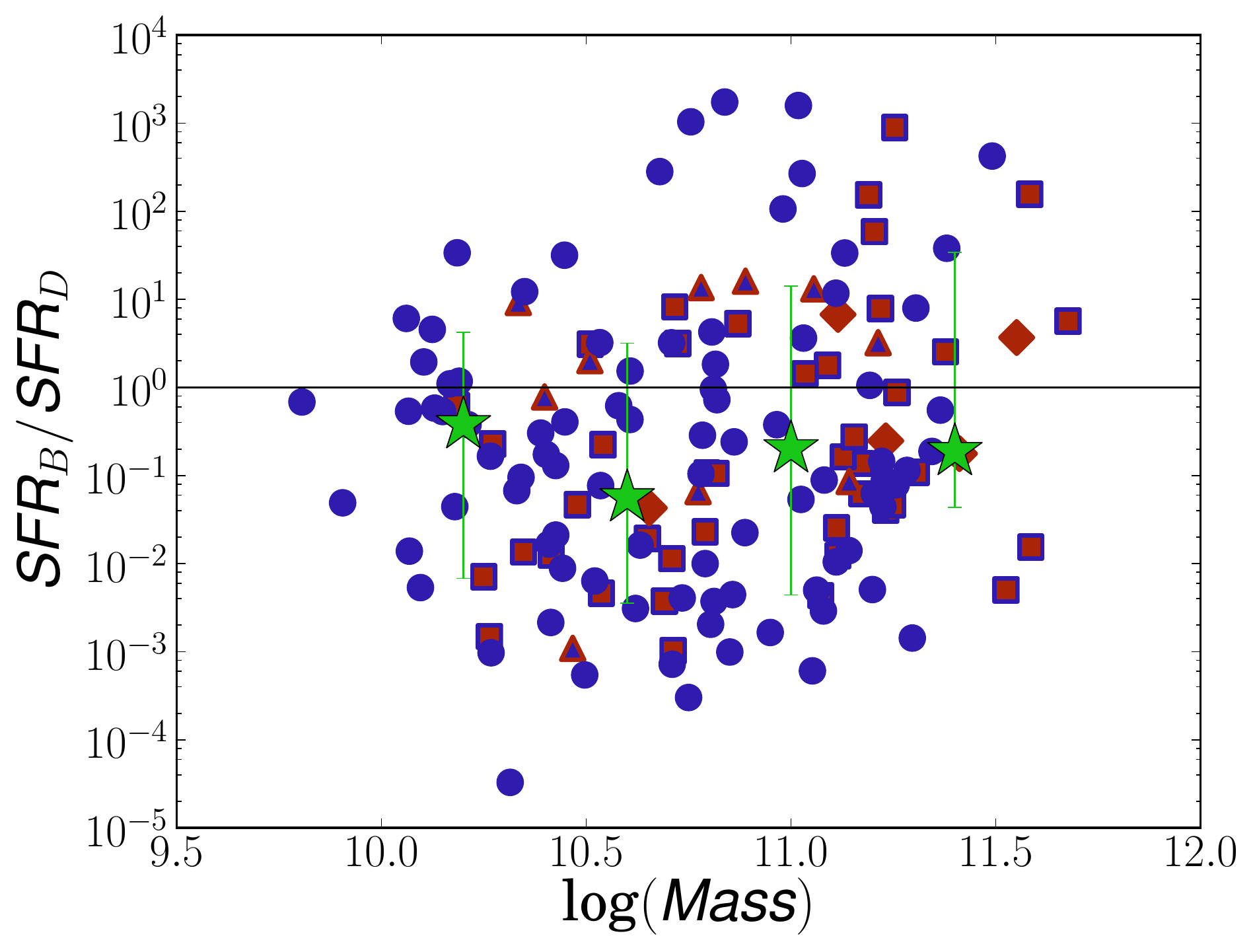}
  \caption{Ratio between SFR of  bulges and discs, as a function of the stellar mass of the galaxy ($M\!ass_B+M\!ass_D$). The inner part of the symbols represents the bulge and the outer disc. Red colour indicates that a certain component is passive while blue indicates that it is star forming. Blue circles are galaxies in which both the bulge and the disc are star forming, red diamond are galaxies that are passive. A square represents a galaxy with a passive bulge and a star-forming disc. Finally, a triangle indicates a star-forming disc and passive bulge. Green stars are the median value in stellar mass bins. The error bars show the dispersion of the data.}\label{fig.ratio_sfr_vs_mass}
\end{figure}

We also examine how the ratio of the star formation in the inner and outer components changes with redshift and stellar mass. The ratio of SFR of bulges and discs is on average lower than unity at all masses (Figure \ref{fig.ratio_sfr_vs_mass}). This implies that the majority of galaxies tend to have higher SFRs in their discs than in their bulges.

\begin{figure*}
  \includegraphics[width=0.49\linewidth]{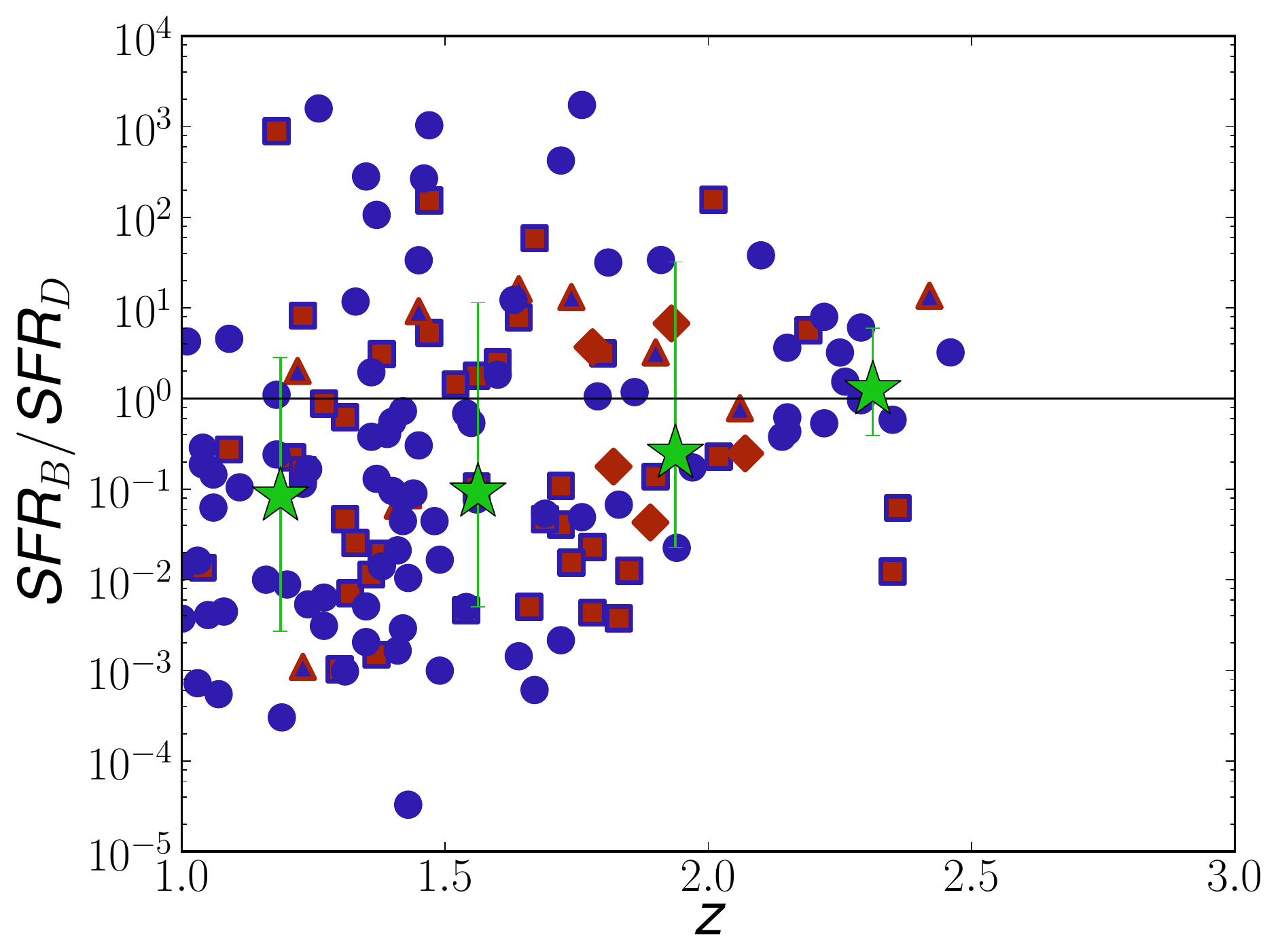}
  \includegraphics[width=0.49\linewidth]{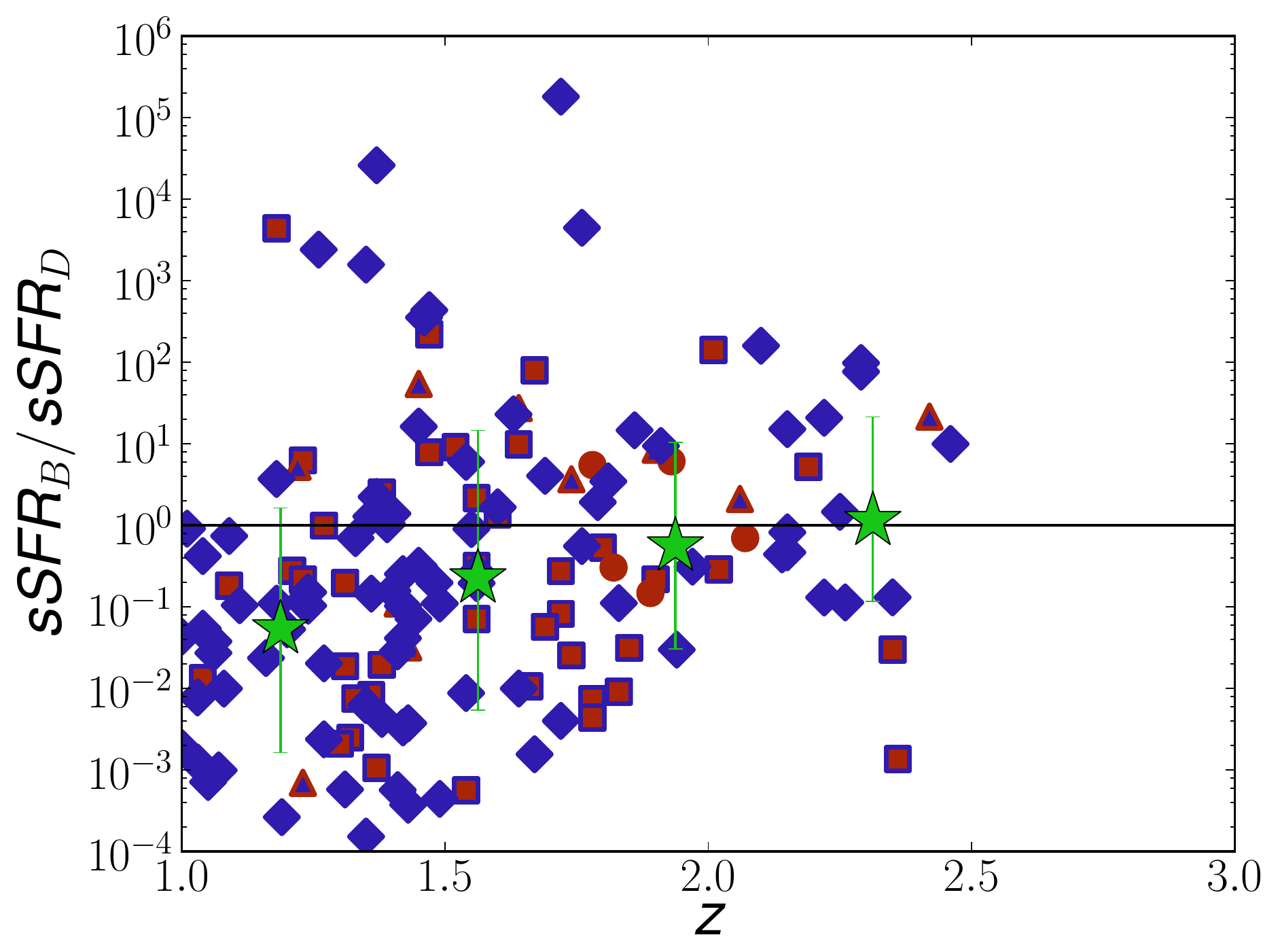}
  \caption{Ratio between SFR of  bulges and discs (left) and ratio between sSFR of bulges and discs (right), as a function of redshift. Colours as in Figure \ref{fig.ratio_sfr_vs_mass}. Green stars are the average over redshift bins. The error bars show the dispersion of the data.}\label{fig.ratio_sfr_vs_z}
\end{figure*}

\begin{figure*}
  \includegraphics[width=0.49\linewidth]{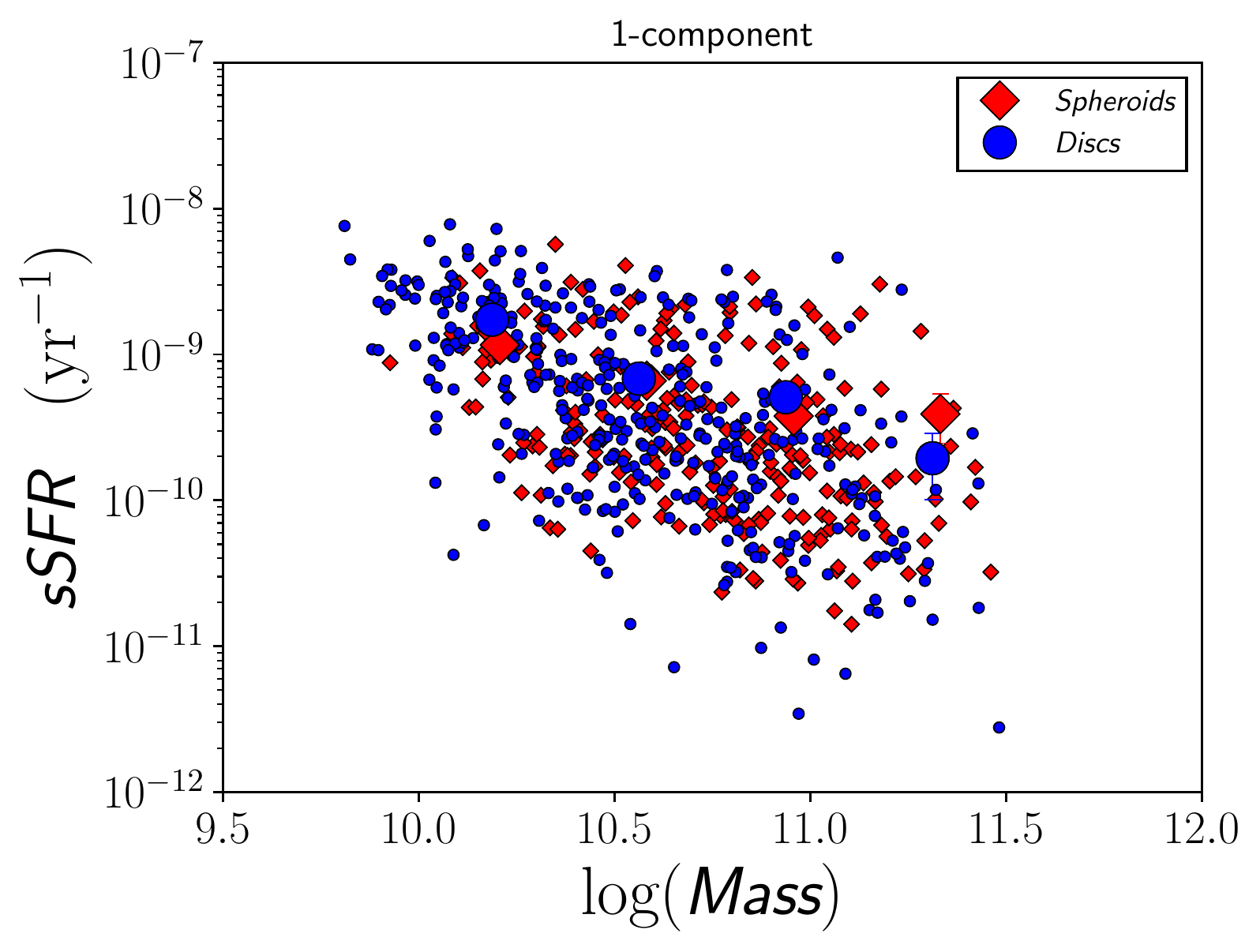}
  \includegraphics[width=0.49\linewidth]{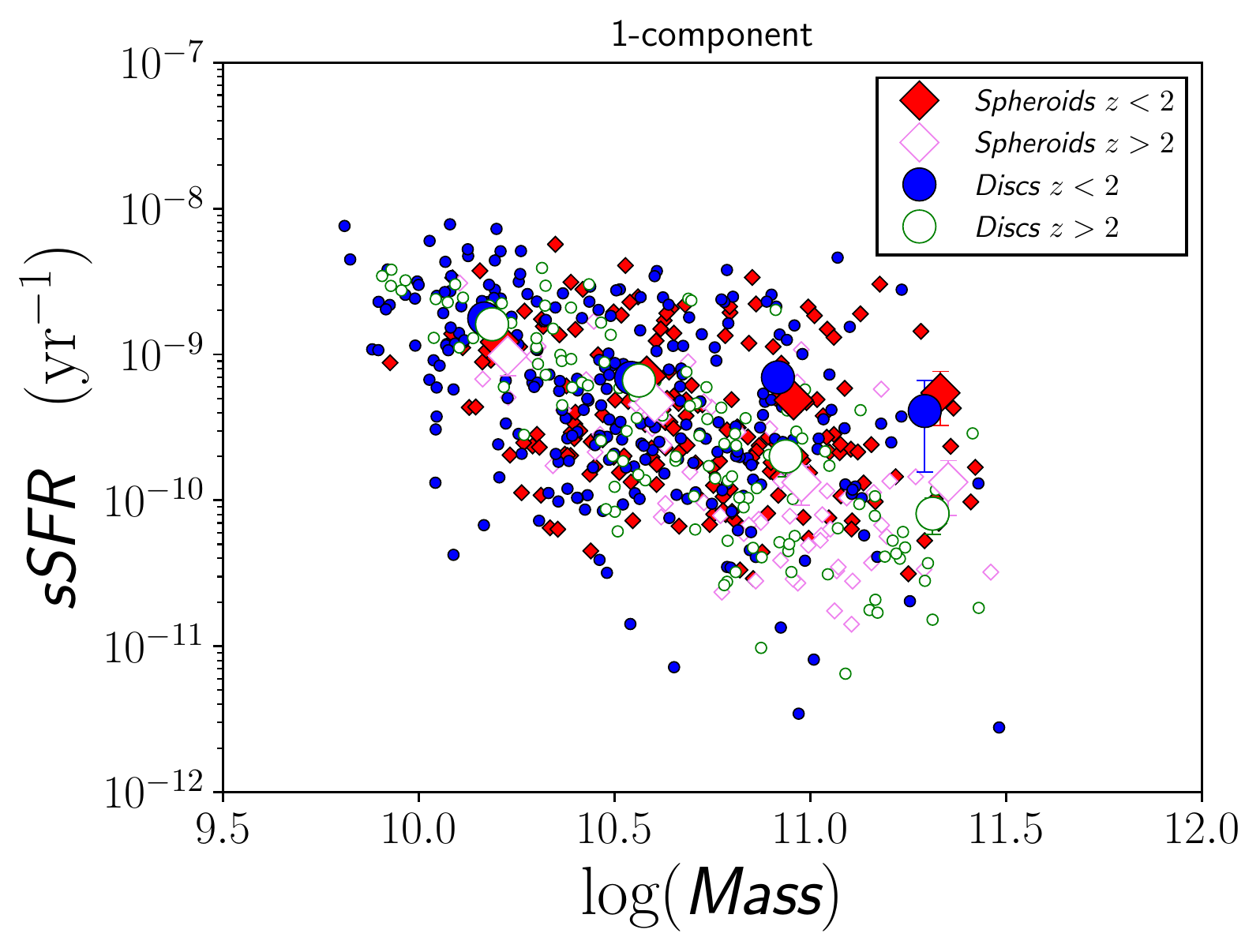}\\
  \includegraphics[width=0.49\linewidth]{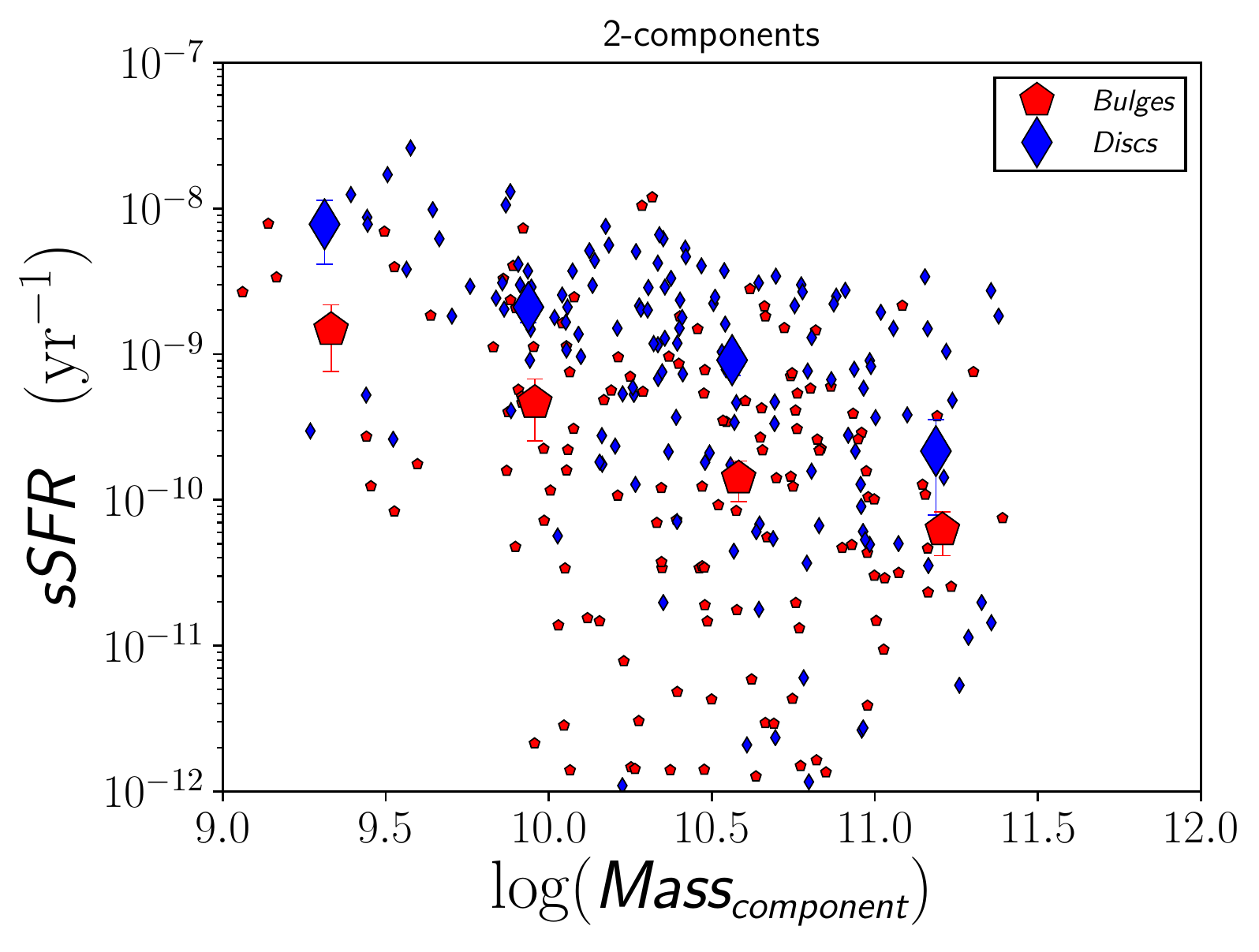}
  \includegraphics[width=0.49\linewidth]{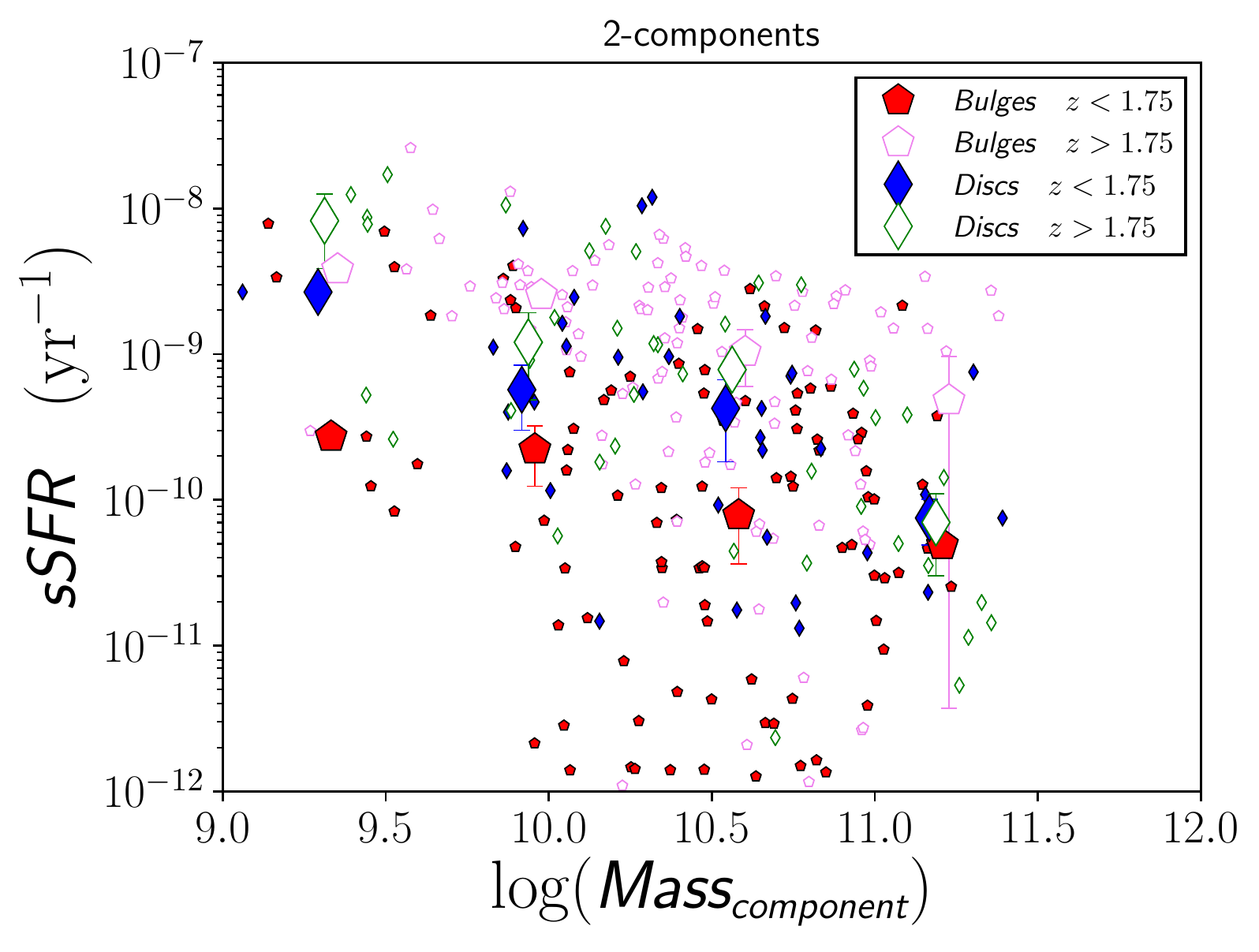}
  \caption{sSFR as a function of stellar mass. On the top left panel we represent the $1$-component galaxies divided by morphology (disc-like galaxies as blue circles and spheroid-like ones as red diamonds). On the right top panel we further split these two types of galaxies into two redshift bins. And analogously for the bulges and discs of the $2$-component objects in the bottom panel. The large symbols are the average sSFR values in each mass bins.}\label{fig.ssfr_vs_mass}
\end{figure*}

Figure \ref{fig.ratio_sfr_vs_z} shows the ratio of SFR with redshift. High redshift bulges have high SFR but as redshift decreases they become more passive while their discs become more star-forming. This is seen as well with the ratio of the specific SFR (sSFR) (Figure \ref{fig.ratio_sfr_vs_z}) thus the effect is not one of a differential in mass between the components. Furthermore, the $1$-component galaxies show a trend of sSFR with stellar mass (Figure \ref{fig.ssfr_vs_mass}, top panels). This trend does not change significantly with morphology. However, we observe a difference with redshift, such that high redshift objects have a stepper relation of sSFR with stellar mass. In particular the sSFR of the most massive galaxies increases with decreasing redshift between $z=3$ to $z=1$. Bulges appear to have a lower sSFR than discs (Figure \ref{fig.ssfr_vs_mass}, bottom panels). However, this is only seen in low redshift bulges, which have very low sSFR values. High redshift bulges have as high of a sSFR than high redshift discs. The sSFR of discs does not change with redshift.

\subsection{Stellar Masses of Components}\label{subsec.mass}

\begin{figure}
  \includegraphics[width=1\linewidth]{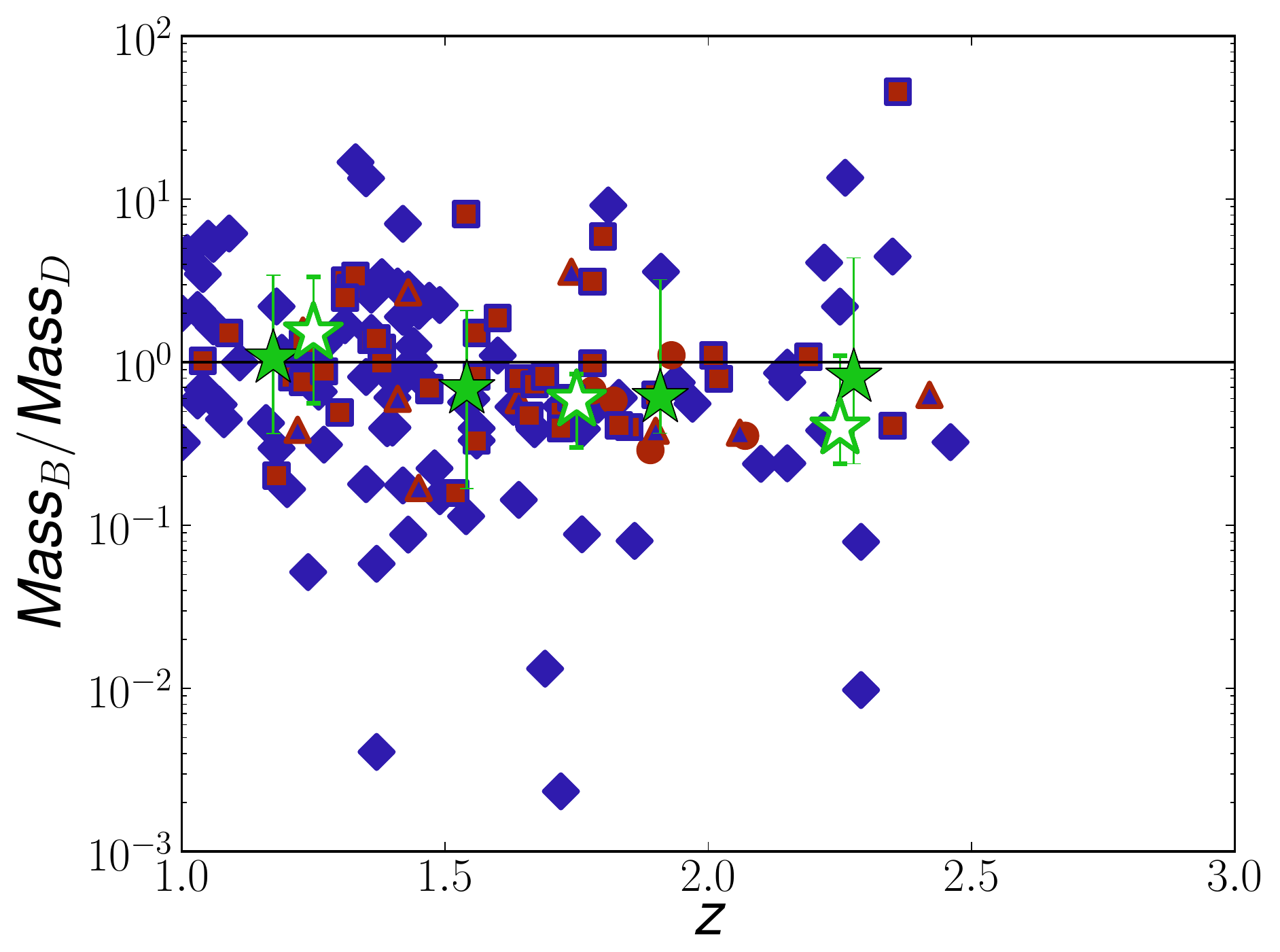}
  \caption{Ratio between stellar mass of bulges and discs as a function of redshift. The green stars indicate the mean values in redshift bins for all $2$-component galaxies. The open star symbols represent the average values for a constant number density selection. Colours are as in Figure \ref{fig.ratio_sfr_vs_mass}. The error bars show the dispersion of the data.}\label{fig.ratio_mass_vs_z}
\end{figure}

\begin{figure}
  \includegraphics[width=1\linewidth]{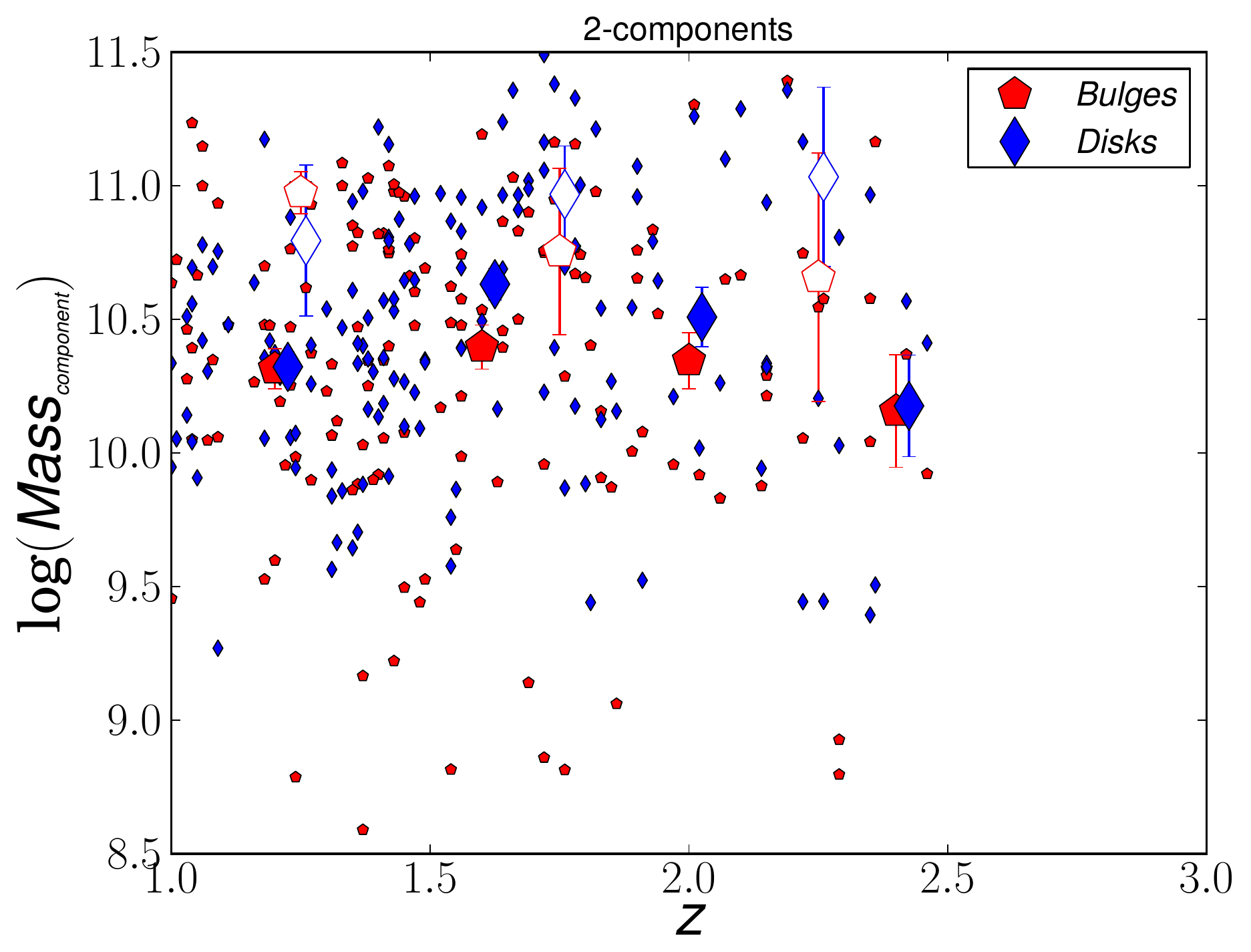}
  \caption{Stellar mass of bulges (red pentagons) and discs (blue diamonds) as a function of redshift. The big filled symbols are the average stellar mass in redshift bins for all the $2$-component galaxies. The big open symbols are the average for a constant number density selection (see text).}\label{fig.mass2_vs_z}
\end{figure}

The ratio of stellar mass between the bulge and the disc is on average constant with time (see Figures \ref{fig.ratio_mass_vs_z} and \ref{fig.mass2_vs_z}). This shows that whilst the masses of both of these components must be changing with time due to star formation, the growth in both goes roughly together. It is easy to see how this occurs for the star-forming outer portions, but the mass assembled in the inner portion, or bulge, is not so direct, as it does not have a constant high SFR. We discuss this later in the discussion section about how this process may occur.

As Figure \ref{fig.mass2_vs_z} shows, there is a relatively good agreement between the masses of the bulges and discs over cosmic time. There is an increase in the average stellar mass, using our selection, between $z = 2.5$ and $z = 2$, and a levelling off of the stellar mass for the two components at later times. If we investigate this as a function of co-moving density, using the derivation for our sample from \cite{Ownsworth16}, we find that from $z=2.5$ to $z=1$ the stellar mass of the bulges increases slightly on average, while there is some decrease in the stellar mass of the disc. This may however be due to other morphological transformations.

\subsection{Presence of AGN}

To identify active galaxies we cross match our galaxy sample with the Chandra X-ray catalogue from Kocevski et al. ($2017$ in prep.). We find that AGNs are more common in spheroid-like galaxies and $2$-component galaxies, with $11\pm2$ and $13\pm3$ per cent of those systems respectively having an AGN, while only $3\pm1$ per cent of disc-like galaxies have an AGN. This shows that the presence of the bulge is correlated with the presence of an AGN, even at high redshift.

\begin{figure*}
  \includegraphics[width=0.49\linewidth]{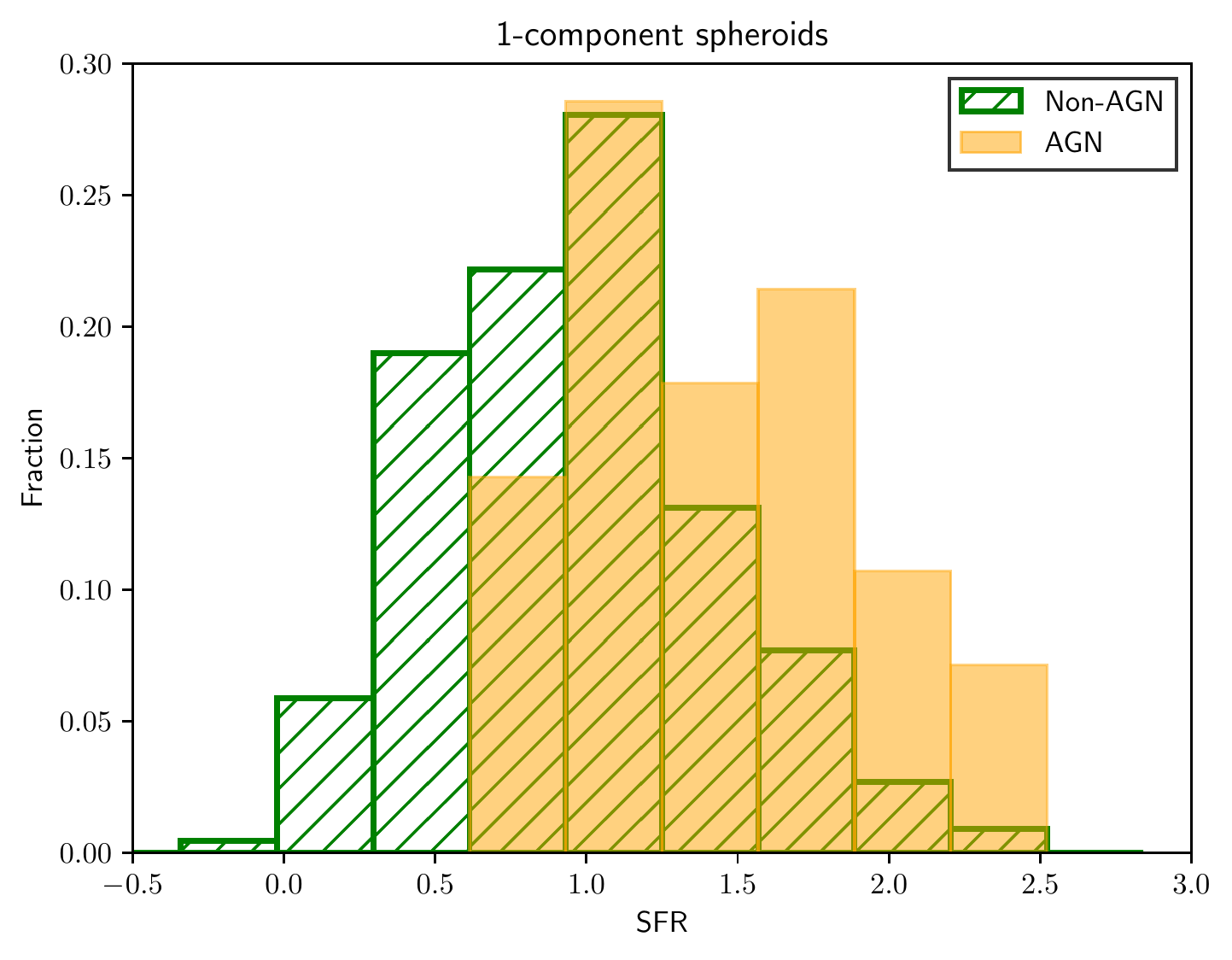}
  \includegraphics[width=0.49\linewidth]{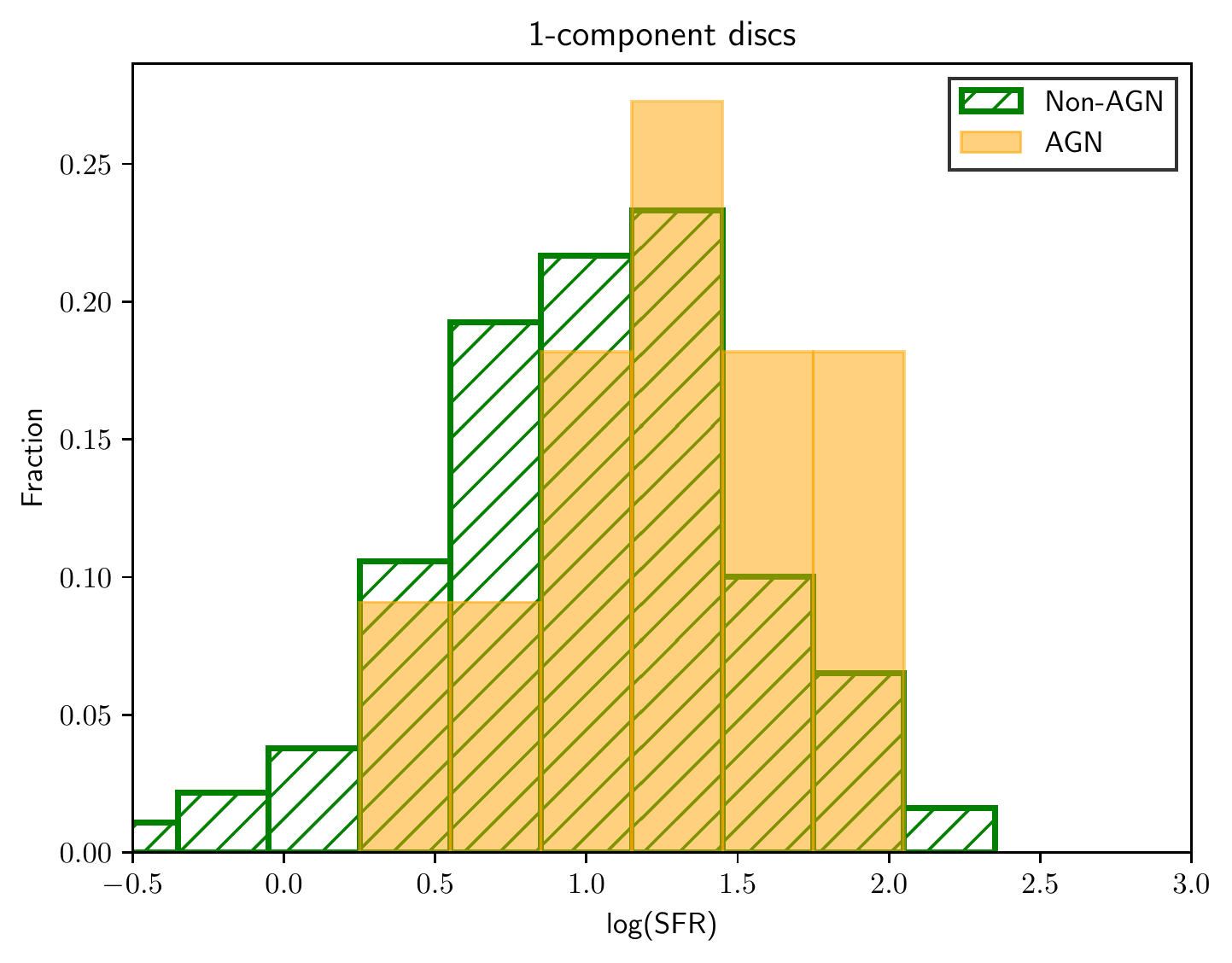}\\
  \includegraphics[width=0.49\linewidth]{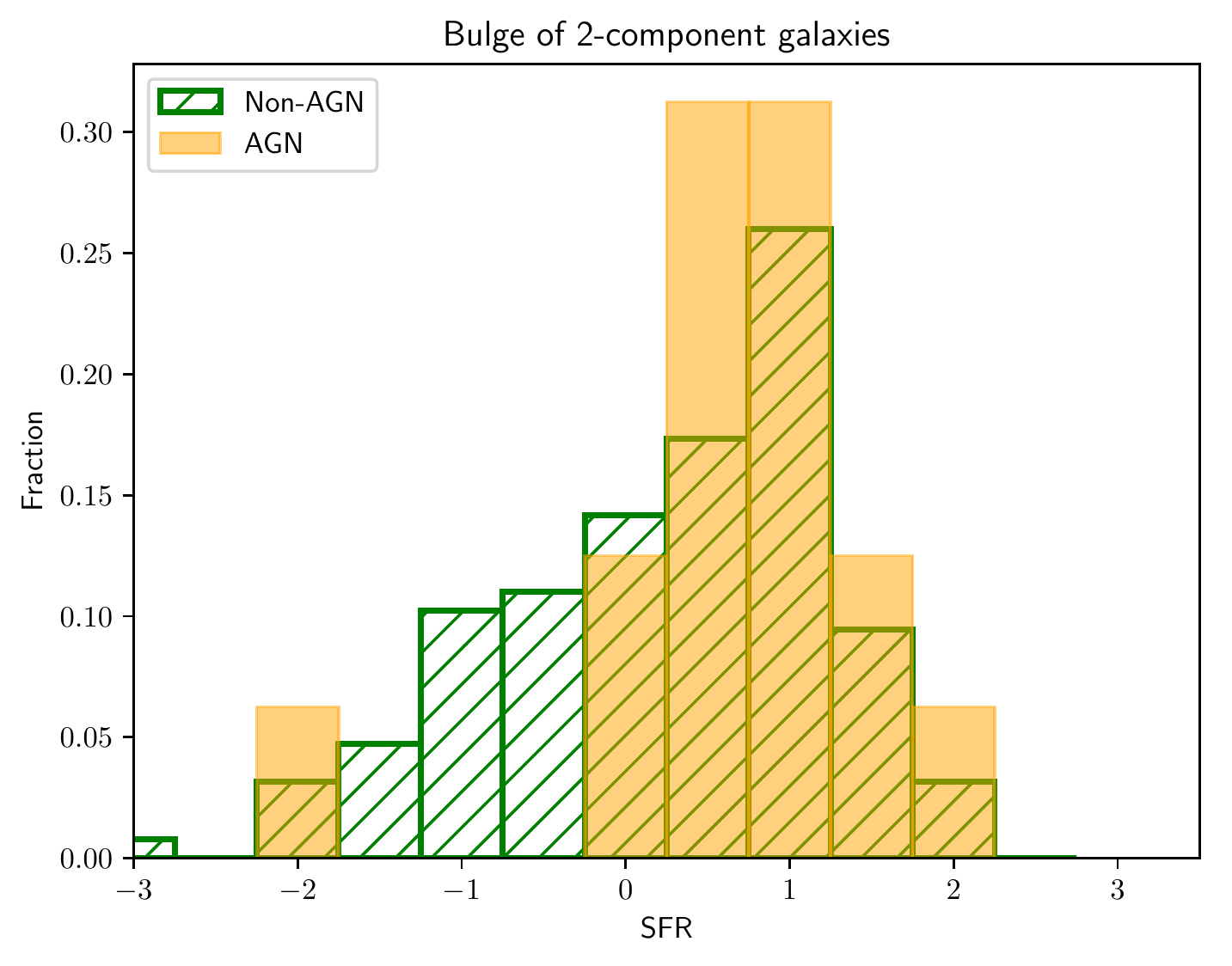}
  \includegraphics[width=0.49\linewidth]{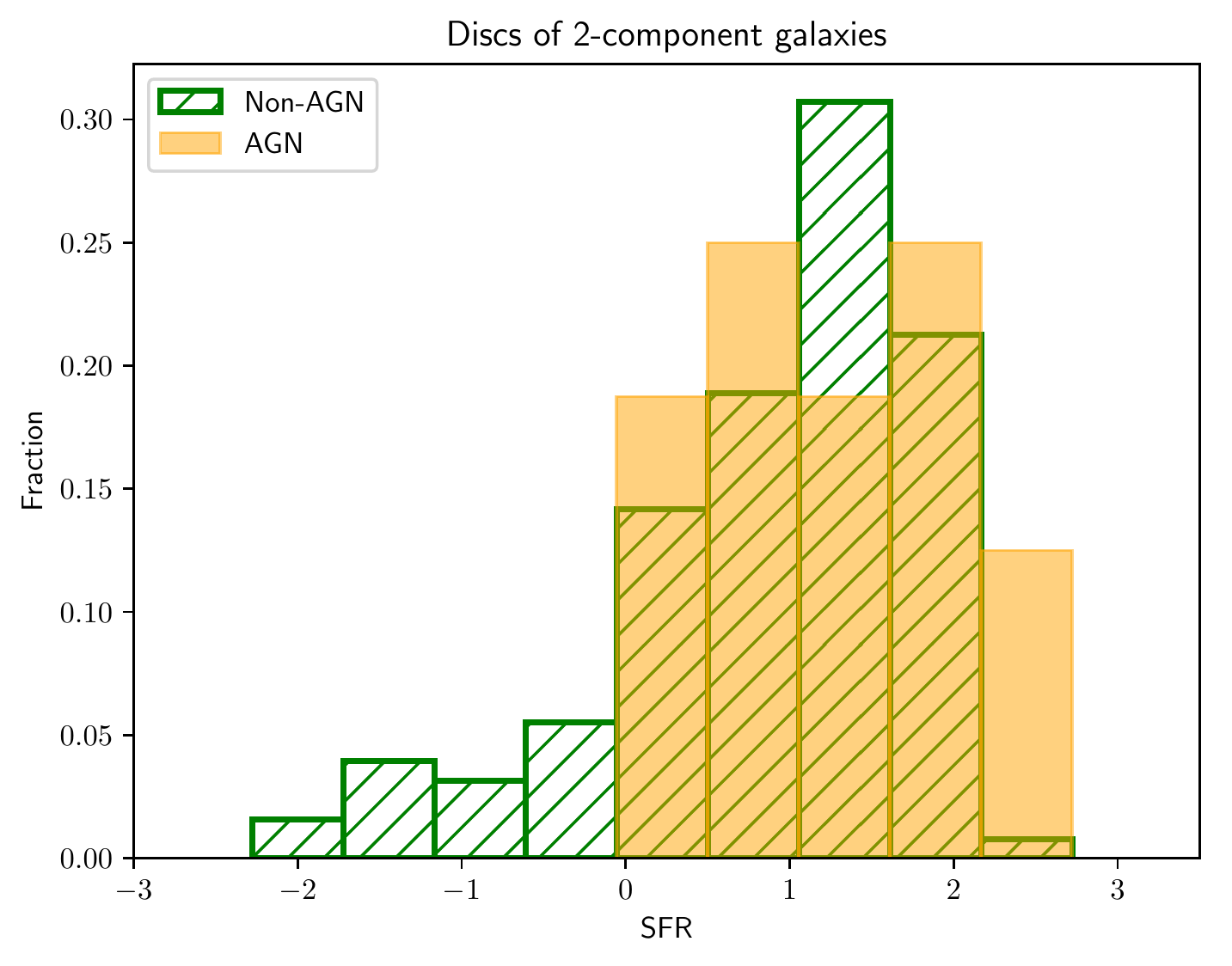}
  \caption{Comparison between the SFR of galaxies with and without the presence of AGN. The top panels show the distribution of $\log (SF\!R)$ for the $1$-components spheroids (left) and discs (right), for galaxies with AGN (solid yellow) and without AGN (strip lines green). The analogous for $2$-component galaxies is shown in the bottom panels for bulges (left) and discs (right).}\label{fig.agn}
\end{figure*}

In Figure \ref{fig.agn} we compare the SFR of galaxies which have an AGN to those that do not have AGN activity. On the top panel we see that $1$-component galaxies with AGN have higher SFR. The difference is stronger for the spheroids. In the bottom panel we show the SFR of the bulge and the disc components for galaxies with and without AGN. We again observe that galaxies with AGN have higher rates of star formation in both components. 

It is possible that the higher SFR observed in is due to the AGN contaminating the host galaxy light \citep{Nandra07,Simmons08}, as AGNs are generally bright in the ultraviolet, which could affect the determination of the SFR from the $UV$ slope. However, in the $2$-component galaxies we observe that the SFR of AGN is higher both in the inner and outer component, although it is expected that only the inner component would be affected by the AGN contamination. 

\section{Discussion}\label{sec.discussion}

In this paper we investigate the properties of the inner and outer portions of galaxies to determine how their formation occurred. To do this we investigate the technical aspects of performing this analysis, using simulations and determining that the SFRs and the stellar masses of both the inner and outer components can be retrieved by assuming that the inner to outer light ratios and properties are similar at the two ends of our ViJH fitting (see Appendix \ref{appendix}). We show that, when examining the bulge to disc ratio for nearby galaxies, this approach is valid, i.e.\, that the bulge to disc ratio does not change significantly with wavelength. This is likely even more true at higher redshifts where the visual morphologies of galaxies and their measured structures do not change as much as for nearby galaxies \citep[e.g.,][]{Conselice00, TaylorMager07}. 

We then decompose our galaxies into inner and outer portions and derive the rest-frame colours, stellar masses, and SFRs for these components. We show that, when we compare the sum of the SFR and stellar masses for the two components to the one component measurement, we obtain a strong correlation. This shows that our method of decomposition is reliable and consistent (see also \cite{Margalef16}). 

When we examine the location of our systems in a $UV\!J$ diagram we find that the vast majority of discs, and around $\sfrac{2}{3}$ of bulges fall into the star-forming region of the $UV\!J$ space. Therefore, a higher fraction of the bulges are more passive. However, what we do find is that many of these star-forming inner components are found in the $UV\!J$ part of the diagram occupied by dusty star-forming systems. Thus, whilst these inner components, or forming bulges, are red, they are not necessarily old, as a significant fraction at $z > 1$ are found in the dusty star-forming region of $UV\!J$ space. Nonetheless, the SFR and sSFR of these bulges are not as high as for the disc or outer components.

We find also that massive ($\log M_{\ast}> 11$) galaxies and galaxy components establish their rest-frame colour by $z=3$ and are the reddest objects amongst our sample. Besides, these objects also do not show evolution in colour with redshift, remaining red and passive.

Moreover, we find a significant difference between the location in $UV\!J$ space of best-fit single profiles \citep[see][]{Margalef16} for systems that are disc-like and those that are spheroid-like. The disc-like objects have a similar fraction of star-forming galaxies to the $2$-component systems, at $78\pm6$ per cent, whereas those systems that are more spheroid-like have a star formation fraction of only $43\pm5$ per cent. This shows that the $2$-component galaxies have a similar broad SFH to $1$-component disc-like galaxies.

We also find that on average the ratio of the stellar mass of the bulge and the disc remains constant with redshift, while the ratio of the SFR decreases with decreasing redshift (Figure \ref{fig.ratio_sfr_vs_z}). Besides, discs become more star-forming than bulges. This supports the idea that galaxies form inside out \citep{Dullo13,Graham15, delaRosa16}, possibly with mass accreting into the disc inducing its star formation. This is consistent with the gas accretion model for how star formation is triggered in these galaxies \citep{Dekel09,Conselice13}. The majority of the star formation within these galaxies is occurring within the discs, precisely at the location where we would expect it to be.

Finally, we find that the ratio of the stellar mass in inner `bulges' and outer `discs' is similar over cosmic time.  That is, it appears that the amount of stellar mass in both inner and outer components is at roughly the same ratio throughout $1<z<3$ when both components are forming. Because there is a higher star formation rate in the outer portion, we would expect the mass of this outer `disc-like' component to grow faster than the inner component where the star formation rate is lower.  However, we do not see this. Because of this, there then must be a conversion, or transfer of stellar mass from the disc  (or outer component) into the bulge (or inner component). We can do a simple calculation to show that this indeed must be the case. We calculate how much extra mass is needed in the inner components when we investigate the change in the stellar mass for both of the inner and outer components for these galaxies. We do this in terms of a constant co-moving volume, so as to avoid issues with new galaxies entering the sample, and to obtain a clean as possible connection between different redshifts \citep[e.g.,][]{Mundy15}. Figure \ref{fig.mass2_vs_z} shows this evolution where we find that the mass of the inner components grows by a factor of $0.3$dex, whilst the outer components decrease in stellar mass by a similar amount. This shows that around half of the mass in the disc components must be transferred into the bulge on average over this time period. This suggests that bulge formation, for an average disc+bulge system, is done in a secular manner rather than simply though a rapid and early formation.

To advance in this field and understand the formation of bulges and discs we need high resolution data at higher redshifts ($z>3$) future telescopes such as the \textit{James Well Space Telescope} (\textit{JWST}) will provide this and allow us to probe the first epoch of disc and bulge formation within the Universe.

\section*{Acknowledgements}

We thank the CANDELS team for their support in making this paper possible, as well as STFC and the University of Nottingham for financial support.

\bibliographystyle{mnras}
\bibliography{references}
\clearpage

\begin{appendices}
\section{Appendix} \label{appendix}

The photometry obtained in this work consists of the four \textit{HST} bands (two visible bands, denoted $V$ and $i$, and two near-infrared ones, denoted $J$ and $H$). We investigate the effects that using only these four bands have in the results of the SED fitting. To this end, we use the galaxy photometry from \cite{Galametz13} for the same galaxies in our sample. 

\begin{figure*}
  \includegraphics[width=0.49\linewidth]{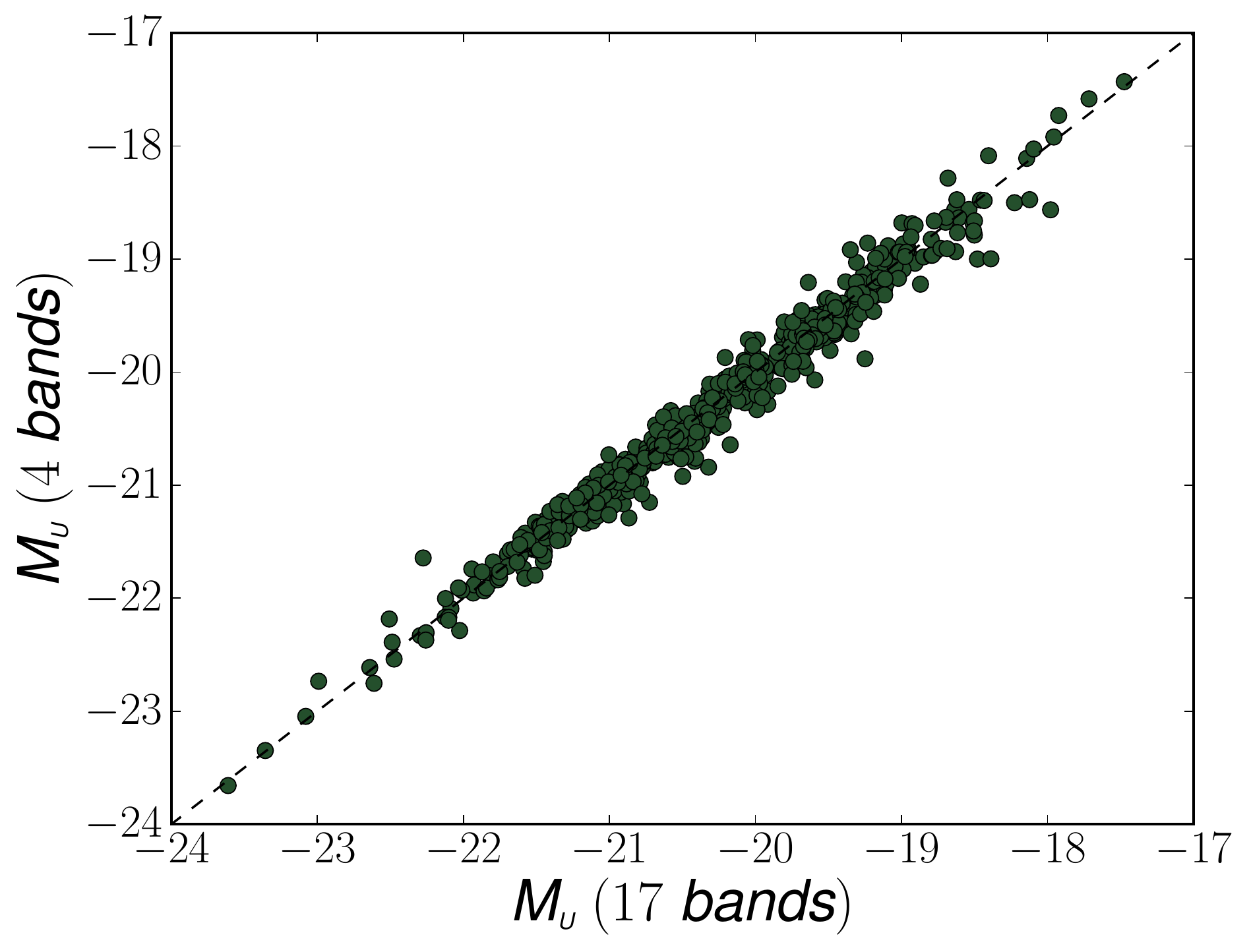}
  \includegraphics[width=0.49\linewidth]{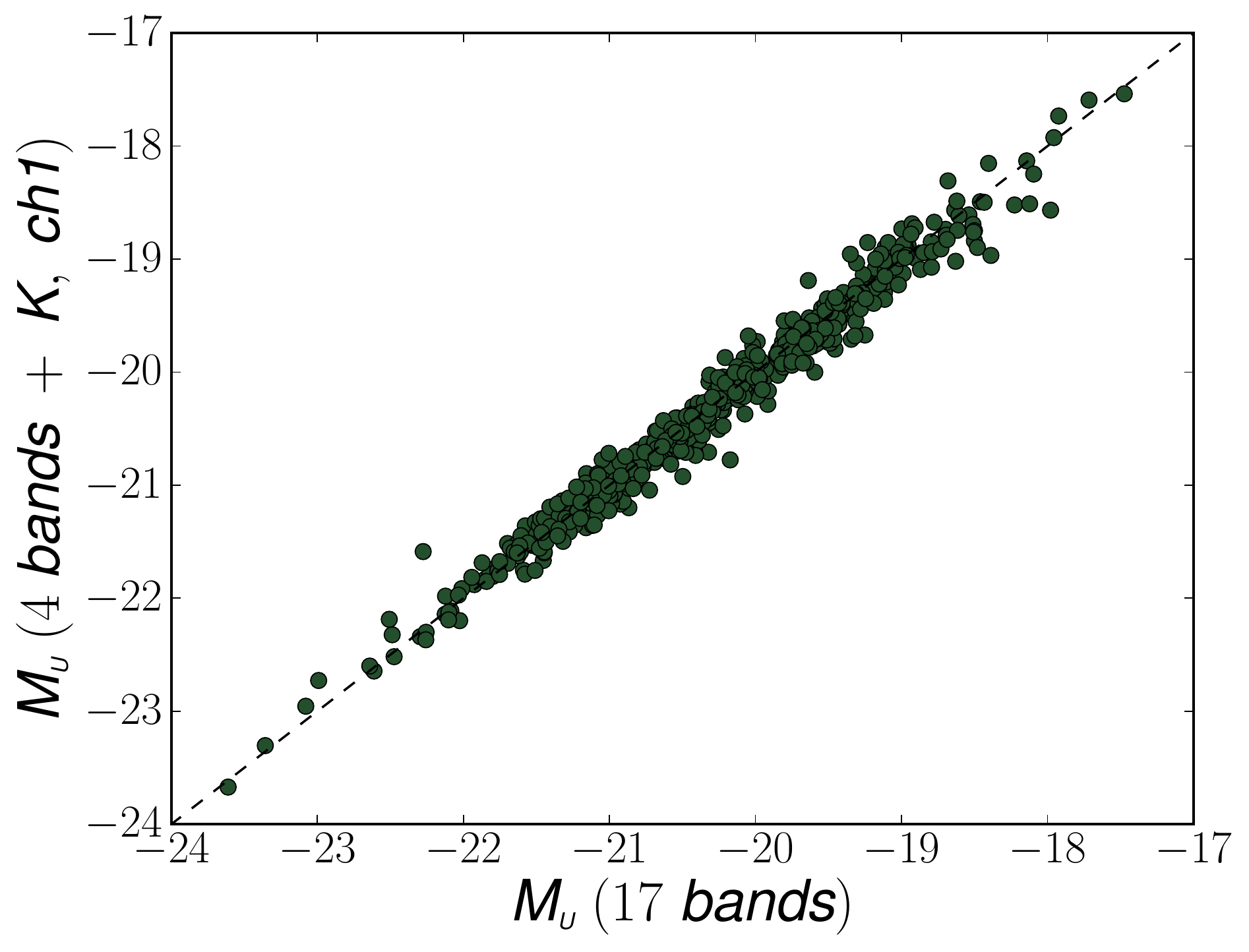}\\
  \includegraphics[width=0.49\linewidth]{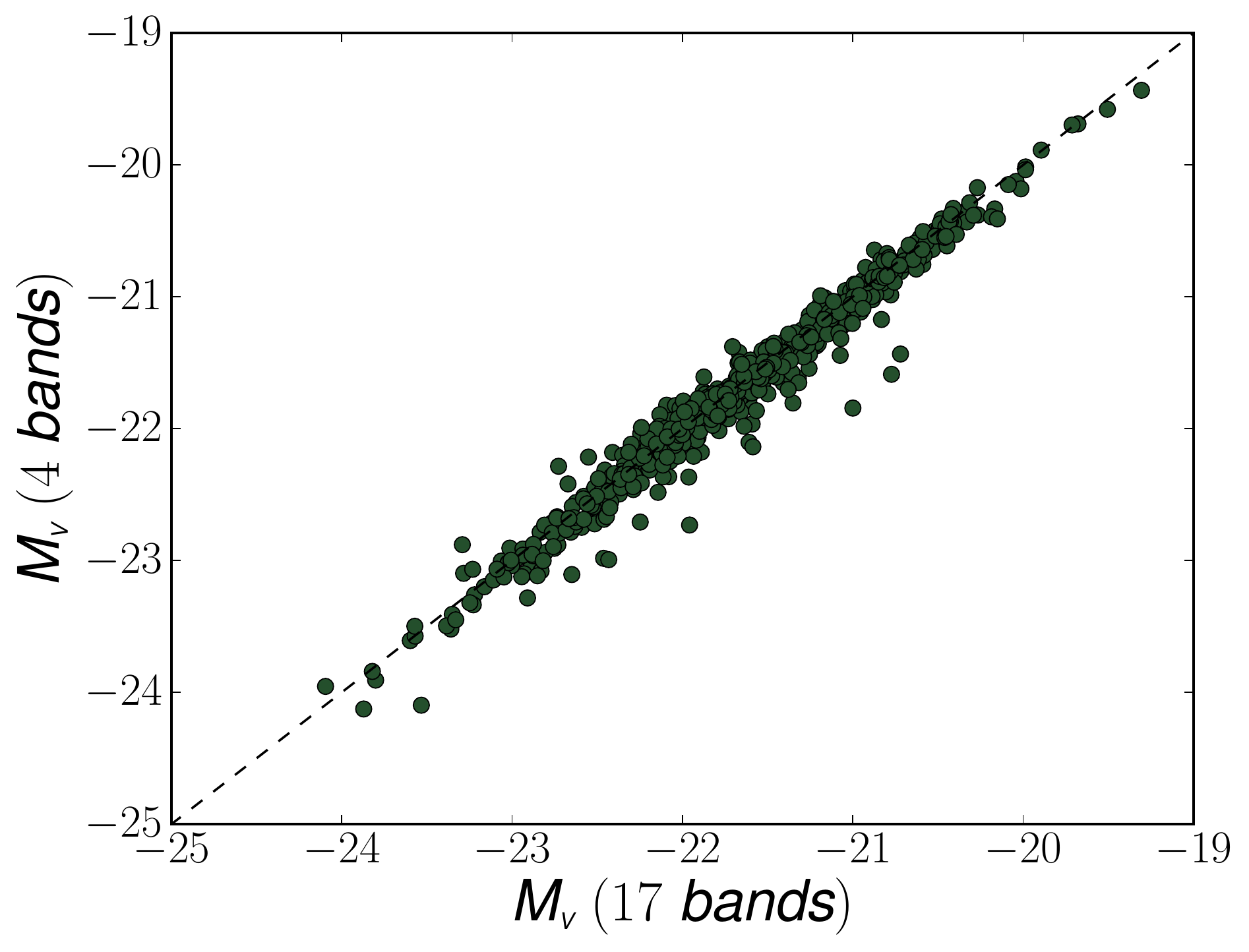}
  \includegraphics[width=0.49\linewidth]{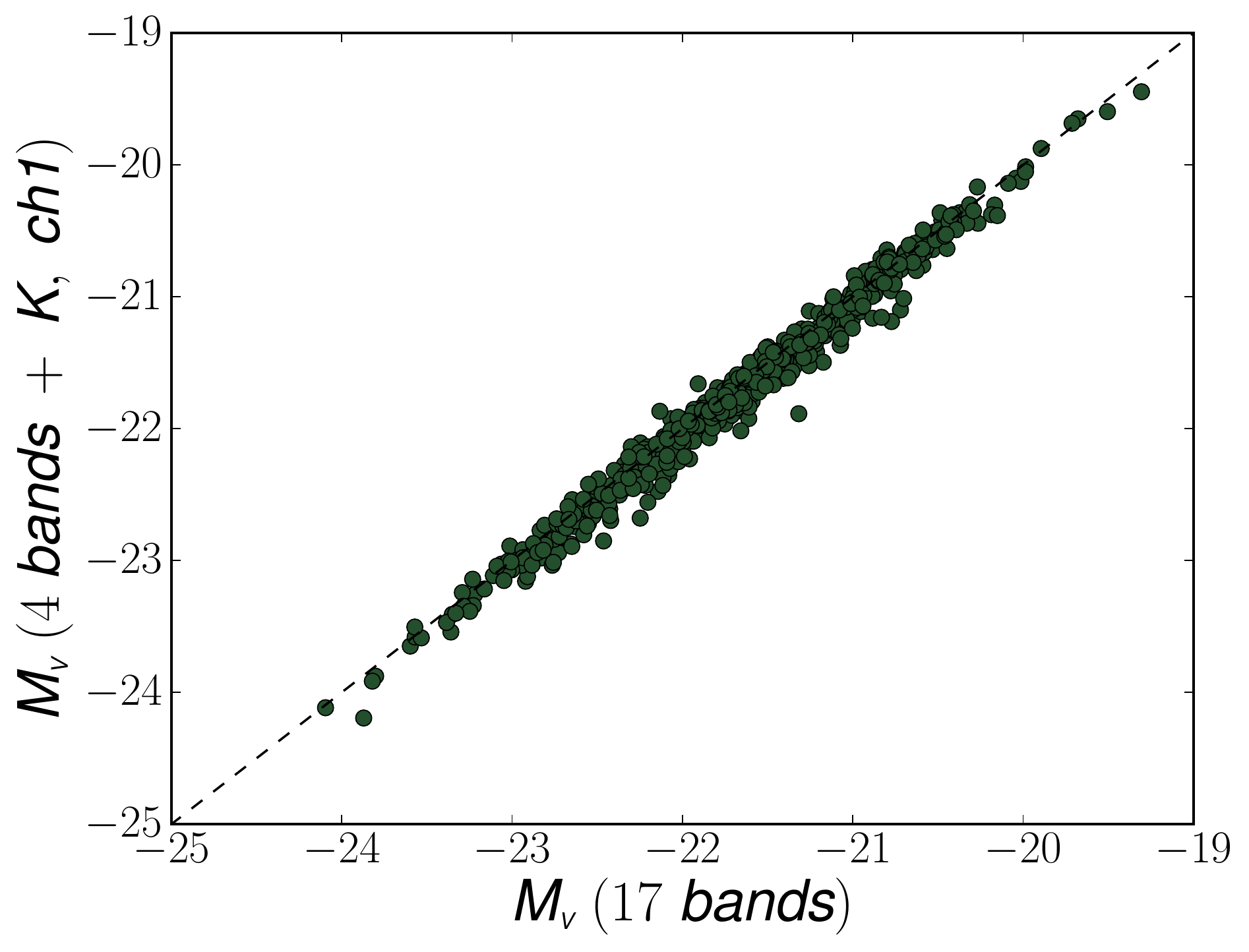}\\
  \includegraphics[width=0.49\linewidth]{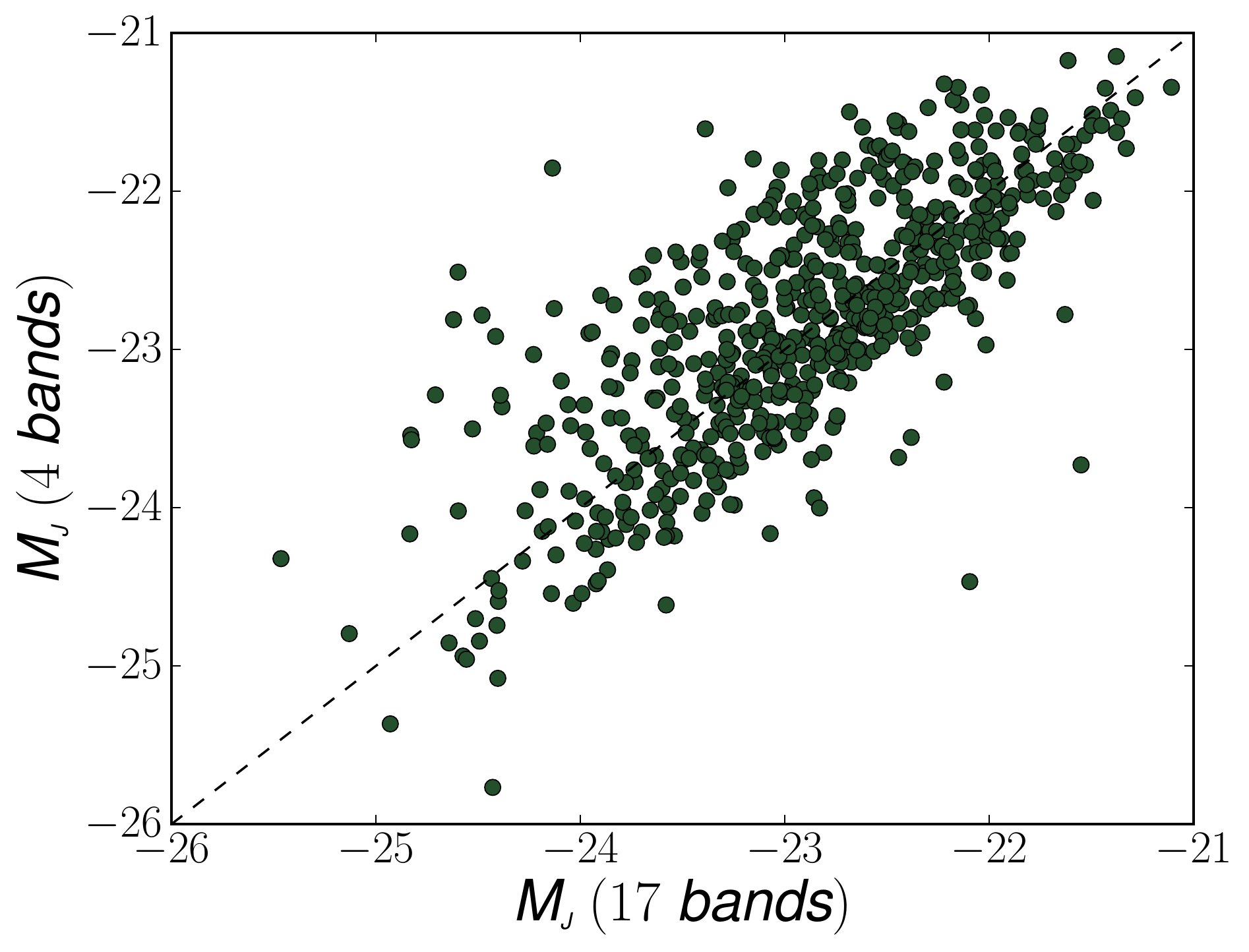}
  \includegraphics[width=0.49\linewidth]{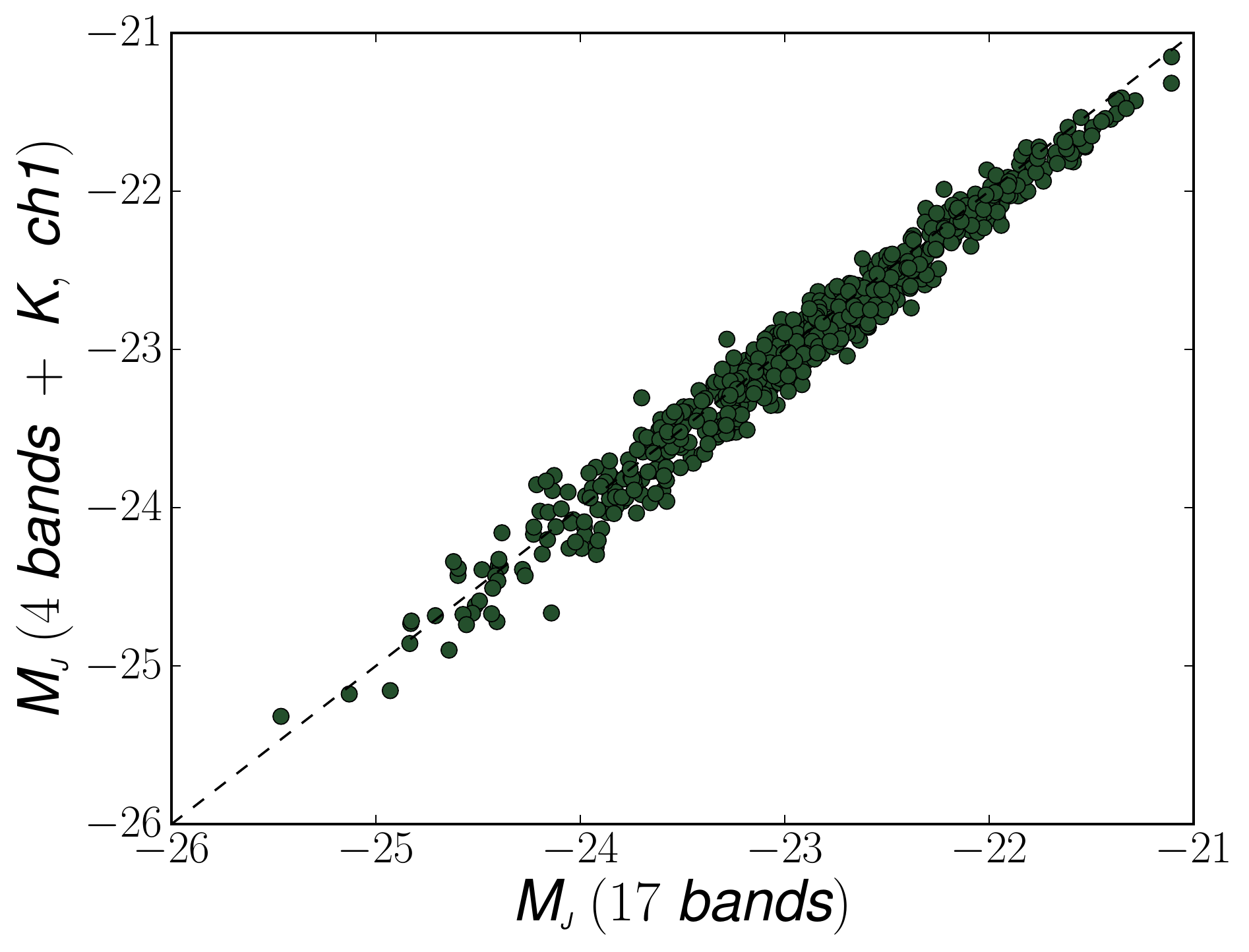}
  \caption{On the left panels, we compare the computed rest-frame magnitudes $U$ (top), $V$ (middle) and $J$ (bottom) obtained using $4$ \textit{HST} bands ($H$, $J$, $i$, $V$) and $17$ bands from \protect\cite{Galametz13}. On the right panels we show how the rest-frame magnitudes change by adding $K$ and $ch1$ band to the $4$ \textit{HST} bands, in comparison with those obtained from the $17$ bands photometry.}\label{fig.4_vs_17}
\end{figure*}

The $17$ bands used for this photometry are described in \S \ref{sec.data}, and include the $4$ \textit{HST} bands used in this work. We discuss the comparison in derived properties between using the $4$ \textit{HST} bands, and using the photometric catalogue from \cite{Galametz13}. We show the results of this comparison on the left side of Figure \ref{fig.4_vs_17}, where we observe that, due to not having rest-frame IR fluxes from Spitzer, the $J$ rest-frame colour is not well recovered, however, $U$ and $V$ are. We then show, on the right panels of Figure \ref{fig.4_vs_17}, that by adding only $K$ and $ch1$ bands, we are able to recover the rest-frame magnitude $J$ as well as when using the $17$ photometric bands.

\begin{figure*}
  \includegraphics[width=0.49\linewidth]{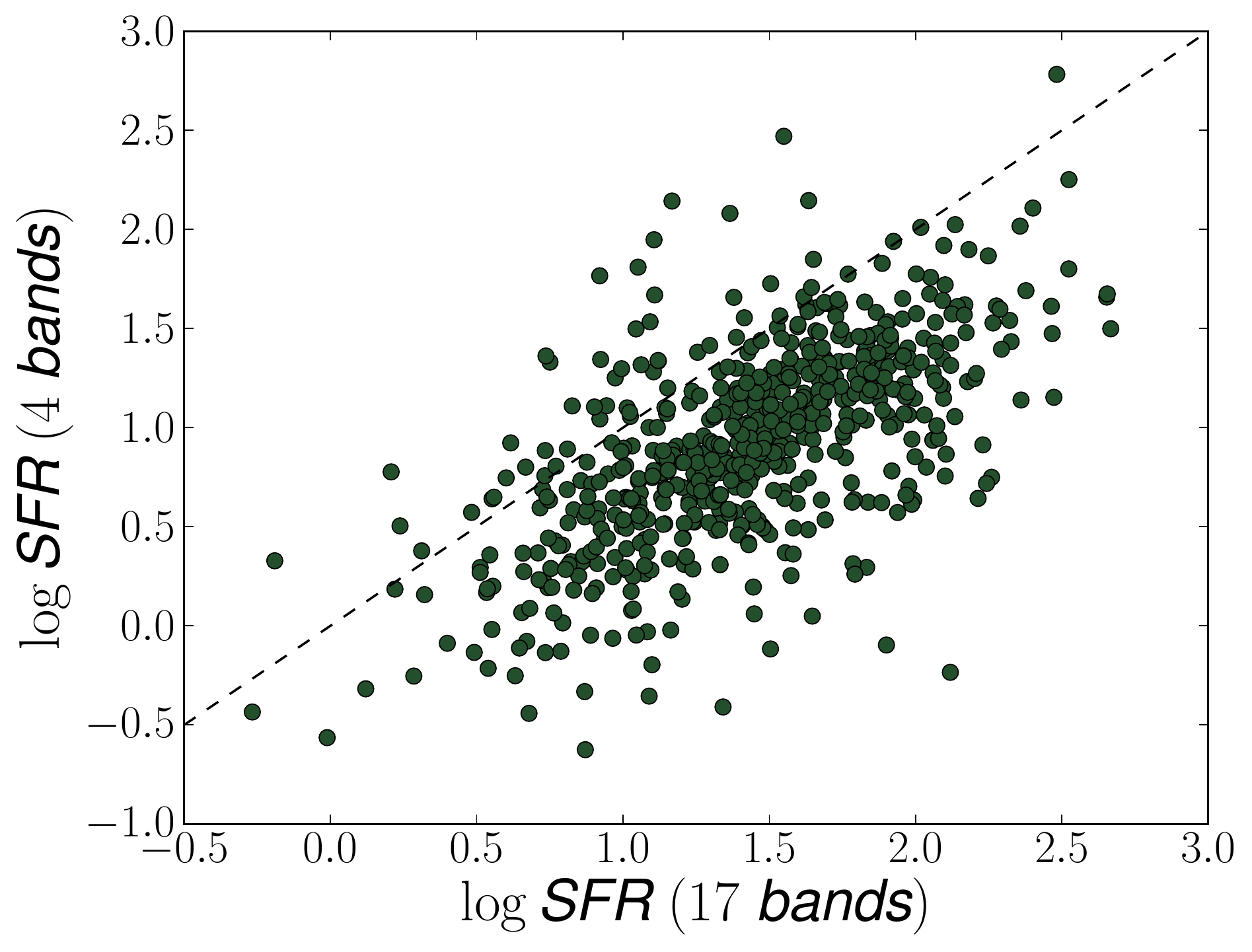}
  \includegraphics[width=0.49\linewidth]{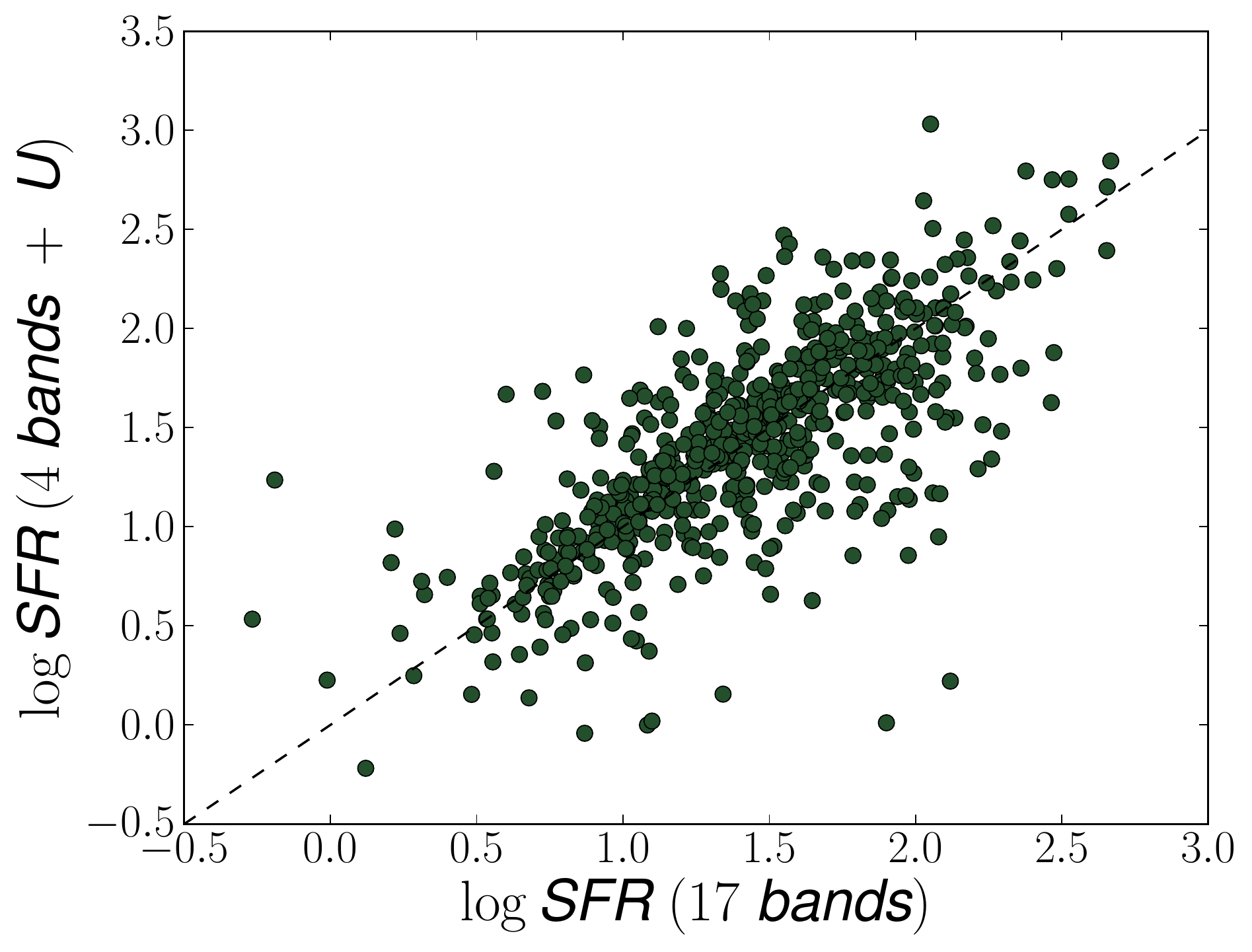}
  \caption{On the left panels, we compare the SFR obtained using $4$ \textit{HST} bands ($H$, $J$, $i$, $V$) and $17$ bands from \protect\cite{Galametz13}. On the right panels we show how the SFR changes by adding the $U$ band to the $4$ \textit{HST} bands, in comparison the those obtained from the $17$ bands photometry.}\label{fig.sfr_vs_sfr}
\end{figure*}

In Figure \ref{fig.sfr_vs_sfr} we show that using just the $4$ \textit{HST} bands, we obtain systematically lower SFRs than using all $17$ bands. This is due to not having photometry in the ultraviolet, which makes it more difficult to constrain the $UV$ slope in the SED fitting. We find that adding the $U$ band to the $4$ \textit{HST} bands we recover the same SFR as with $17$ bands. We also note that by adding $K$ and $ch1$, apart from $U$, does not change the recovered SFRs.

\begin{figure*}
  \includegraphics[width=0.49\linewidth]{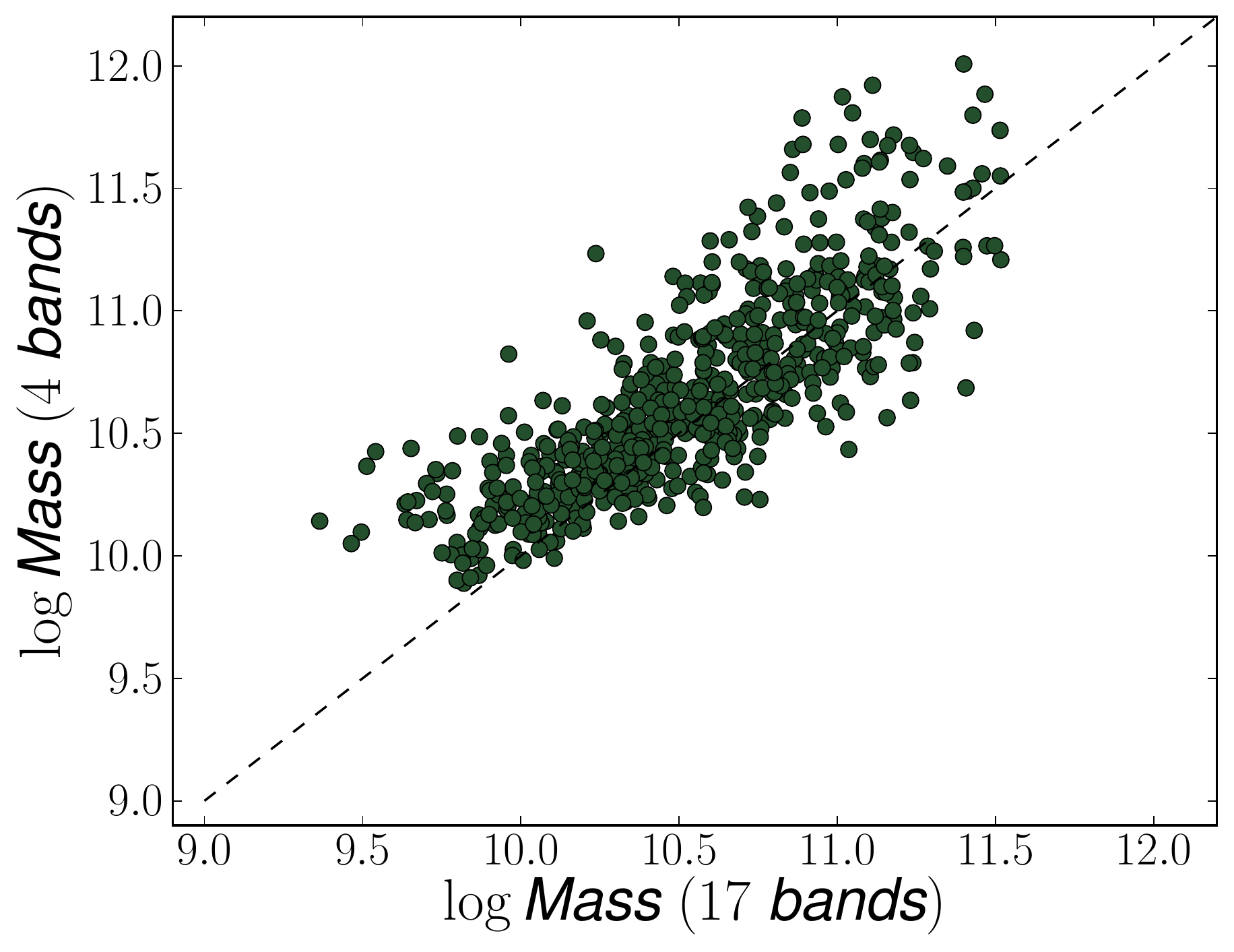}
  \includegraphics[width=0.49\linewidth]{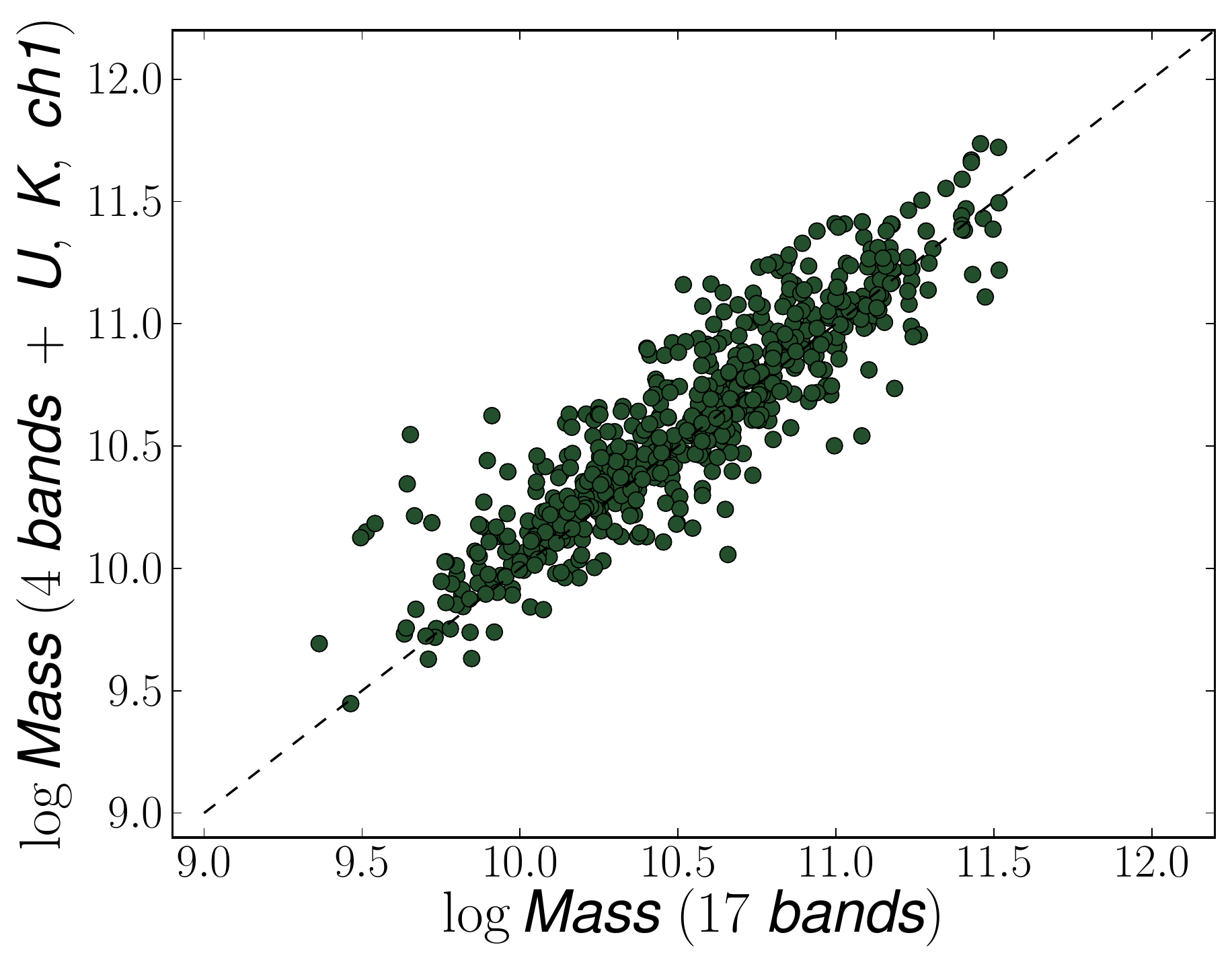}
  \caption{On the left panels, we compare the stellar mass obtained using $4$ \textit{HST} bands ($H$, $J$, $i$, $V$) and $17$ bands from \protect\cite{Galametz13}. On the right panels we show how the stellar mass changes when the $U$, $K$ and $ch1$ bands are added to the $4$ \textit{HST} bands, in comparison to those obtained from the $17$ bands photometry.}\label{fig.mass_vs_mass17}
\end{figure*}

Finally we show in Figure \ref{fig.mass_vs_mass17} that including $U$, $K$ and $ch1$ to the $4$ \textit{HST} bands, we also obtain results of the stellar mass consistent with those obtained using $17$ bands. We conclude that using $UViJHK$ and the Spitzer $ch1$ bands are sufficient to obtain reliable results in terms of rest-frame colours, stellar masses and SFRs from SED fitting. We thus derive what these colours are from the assumptions we make in \S \ref{subsec.SEDs}.

\begin{figure}
  \includegraphics[width=1\linewidth]{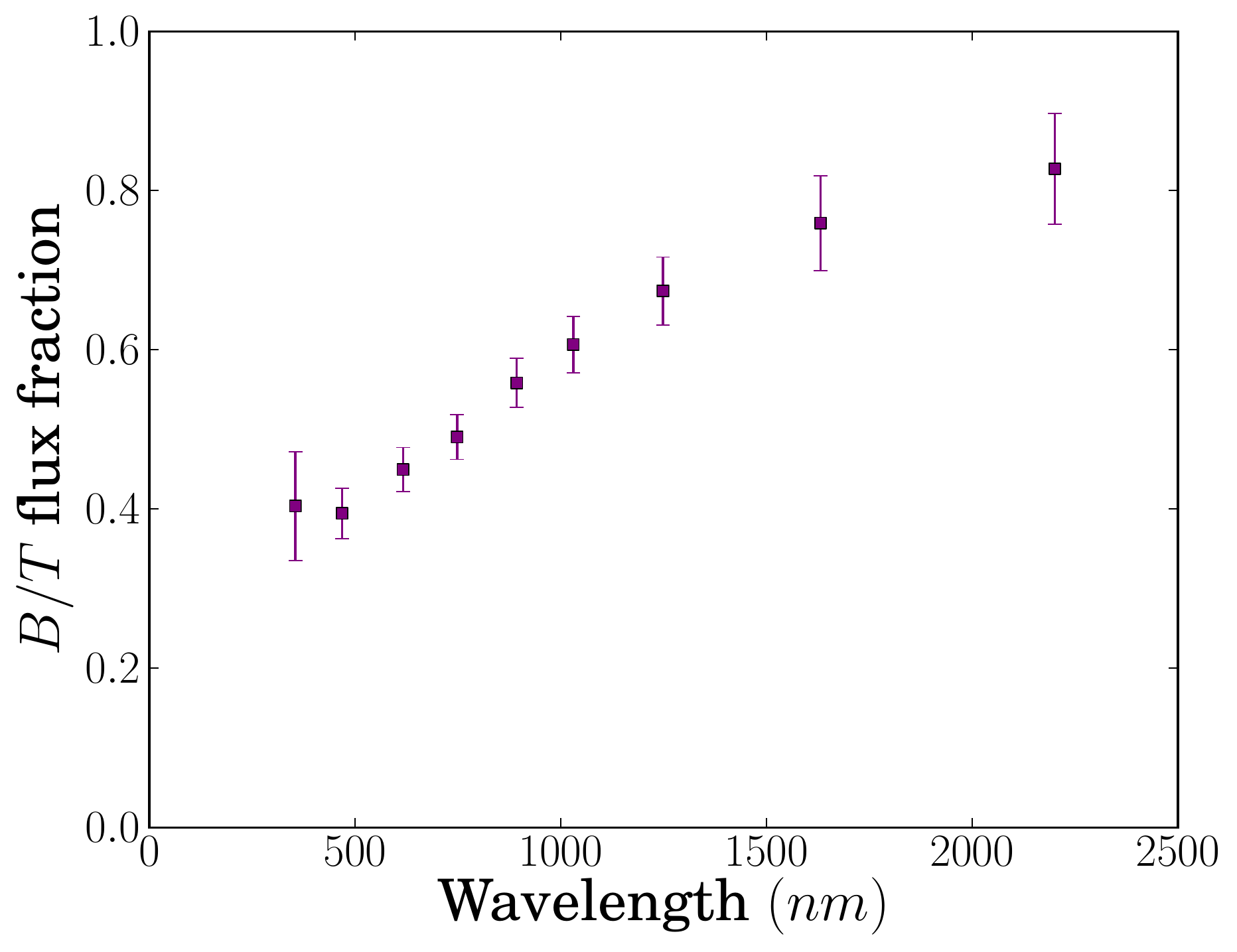}
  \caption{$B/T$ flux ratio as a function of wavelength for a sample of blue nearby galaxies from the GAMA survey.}\label{fig.bt_flux}
\end{figure}

In order to estimate the flux of the components (bulge and disc) in the $K$, $U$ and $ch1$ bands, we first examine a volume limited sample of nearby galaxies ($z<0.3$, $M_{r}<-21.2$) from GAMA \citep{Liske15} for which a bulge to disc decomposition is available. These sample if formed by the blue galaxies containing a bulge and a disc component \citep{Vulcani14,Kennedy16}. Figure \ref{fig.bt_flux} shows how the $B/T$ changes as a function of rest-frame wavelength for this sample of nearby galaxies. 

\begin{figure}
  \includegraphics[width=1\linewidth]{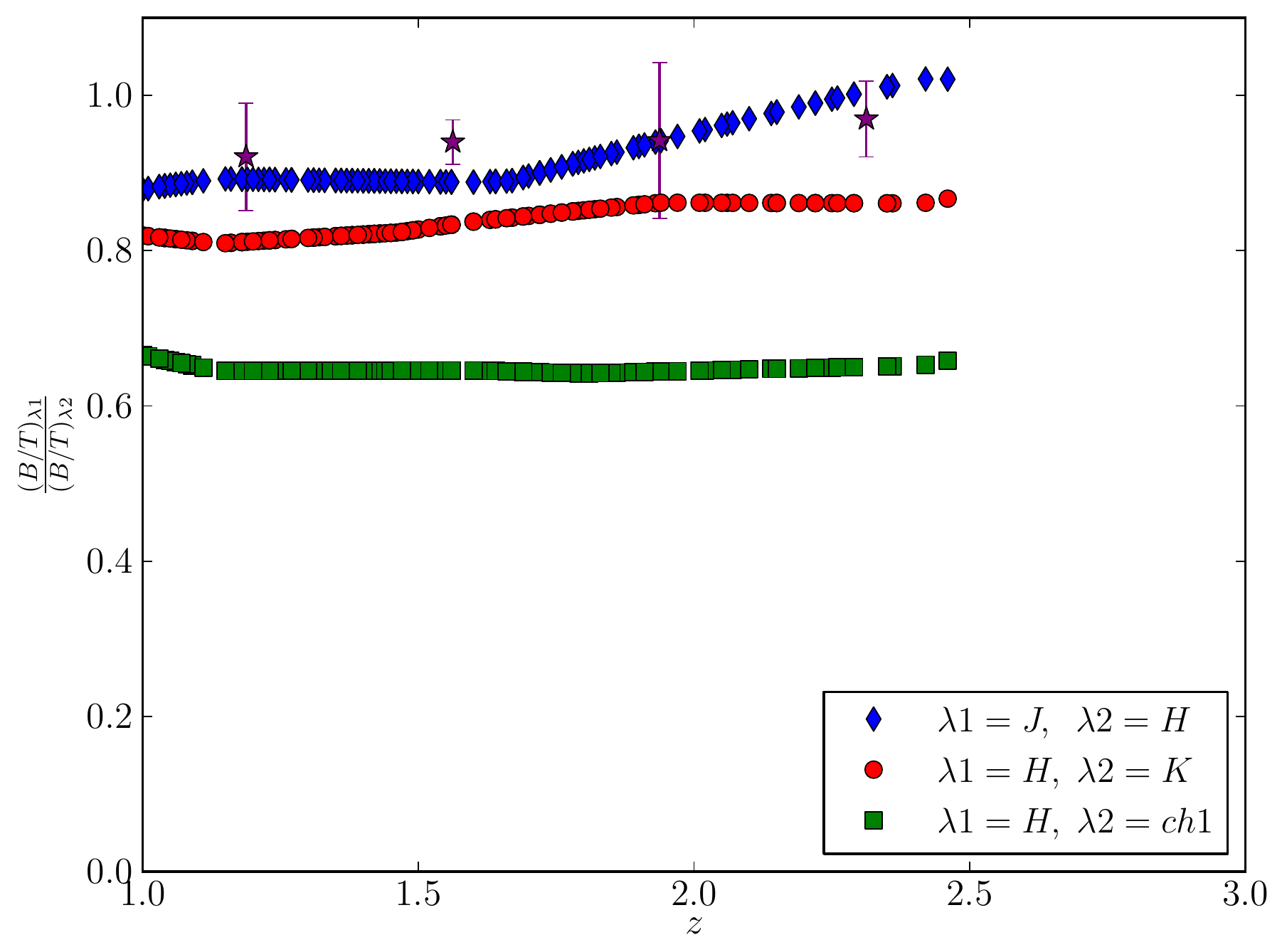}
  \caption{$B/T$ ratio between the rest-frame magnitudes of our $2$-component galaxies corresponding to $H$ and $K$ (red circles), $H$ and $ch1$ (green squares) and $J$ and $H$ (blue diamonds). The $B/T$ ratio is calculated by interpolation of Figure \protect\ref{fig.bt_flux} for the corresponding rest-frame magnitudes in each case. The purple stars are the average ratio between the $B/T$ obtained in the $H$ and $J$ bands from our surface brightness fitting.}\label{fig.ratioBT}
\end{figure}

For each $2$-component galaxy in our high redshift sample we calculate the rest-frame magnitudes that correspond to the observed $H$ and $K$ bands, which will be different for galaxies at different redshifts. We then determine the ratio of $B/T$ between these two rest-frame wavelengths by interpolating from the relation observed in Figure \ref{fig.bt_flux}. We plot these results in Figure \ref{fig.ratioBT} (red circles) and find that this ratio is about $0.85$ for all our galaxies. We can compute the $B/T$ in the $K$ band as follows: 
\begin{equation*}
(B/T)_K=0.85(B/T)_H.
\end{equation*}

\noindent We can therefore calculate the flux of the bulge and disc component in the K-band. Analogously, for $ch1$ we obtain:
\begin{equation*}
(B/T)_{ch1}=0.65(B/T)_H.
\end{equation*}

\noindent To confirm whether the relation of $B/T$ with wavelength from nearby galaxies can be applied to high redshift ones, we compare the ratio of $B/T$ between $H$ and $J$ bands from our surface brightness fittings to those obtained from the low redshift sample from Figure \ref{fig.bt_flux}. We find that in both cases 
\begin{equation*}
(B/T)_J\sim0.9(B/T)_H
\end{equation*}
(blue diamonds and purple stars in Figure \ref{fig.ratioBT}).

Figure \ref{fig.bt_flux} shows that the relation of $B/T$ with wavelength flattens towards the ultraviolet and, therefore, we  assume that the $B/T$ in the $U$ band will be the same as in the $V$ band for all redshifts: 
\begin{equation*}
(B/T)_U=(B/T)_V.
\end{equation*}
    
We use these corrections to obtain photometry for the disc and bulge components in the $U$, $K$ and $ch1$, and use them in addition to the photometry from our $4$ \textit{HST} bands ($V$, $i$, $J$, $H$) to perform SED fittings.

\end{appendices}

\bsp
\label{lastpage}
\end{document}